\documentclass[a4paper,11pt]{article}
\pdfoutput=1
\usepackage[utf8]{inputenc}
\usepackage{amssymb}
\usepackage{amsmath}
\usepackage{amsthm}
\usepackage{amsfonts}
\usepackage[pdftex]{graphicx}
\usepackage{geometry}
\usepackage{braket}
\usepackage{enumerate}
\usepackage{comment}

\usepackage{subcaption}

\allowdisplaybreaks

\usepackage[table,usenames,dvipsnames]{xcolor}

\usepackage{setspace}
\usepackage[in]{fullpage}
\usepackage{hyperref}
\usepackage{array}
\usepackage{cancel}
\usepackage{wrapfig}
\usepackage[font=small,labelfont=bf]{caption}
\usepackage{pifont}

\usepackage[dvipsnames]{xcolor}
\usepackage{mathtools}
\usepackage{todonotes}

\usepackage{color}

\usepackage{tikz}
\usetikzlibrary{arrows,shapes}
\usetikzlibrary{trees}
\usetikzlibrary{matrix,arrows} 
\usetikzlibrary{positioning}
\usetikzlibrary{calc,through}
\usetikzlibrary{decorations.pathreplacing}
\usepackage{pgffor}	

\numberwithin{equation}{section}

\usepackage[english]{babel}

\usepackage{slashed}
\usepackage{caption}

\usepackage[nottoc,notlot,notlof]{tocbibind}
\usepackage[nosort]{cite}   
\usepackage{color}

\usepackage{parskip}
\usepackage[T1]{fontenc}

\usepackage{lmodern}
\usepackage{sectsty}
\usepackage{yfonts}
\allsectionsfont{\boldmath}

\usepackage{physics}
\usepackage{slashed}
\usepackage{makeidx}
\usepackage[vcentermath]{youngtab}
\usepackage{amsthm}
\usepackage{mathtools}

\hypersetup{
	colorlinks=true,
	linktoc=page,
	citecolor=blue,
	linkcolor=blue,
	urlcolor=blue} 

\usepackage{tikz}
\usetikzlibrary{arrows,shapes}
\usetikzlibrary{trees}
\usetikzlibrary{matrix,arrows} 				
\usetikzlibrary{positioning}				
\usetikzlibrary{calc,through}				
\usetikzlibrary{decorations.pathreplacing}  
\usepackage{pgffor}							

\usetikzlibrary{decorations.pathmorphing}	
\usetikzlibrary{decorations.markings}
\tikzset{
   vector/.style={decorate, decoration={snake, amplitude=1pt, segment length=6pt}, draw},
	provector/.style={decorate, decoration={snake,amplitude=2.5pt}, draw},
	antivector/.style={decorate, decoration={snake,amplitude=-2.5pt}, draw},
    fermion/.style={draw=black, postaction={decorate},
        decoration={markings,mark=at position .55 with {\arrow[draw=black]{>}}}},
    fermionbar/.style={draw=black, postaction={decorate},
        decoration={markings,mark=at position .55 with {\arrow[draw=black]{<}}}},
    fermionnoarrow/.style={draw=black},
    gluon/.style={decorate, draw=black,
        decoration={coil,amplitude=4pt, segment length=5pt}},
    scalar/.style={dashed,draw=black, postaction={decorate},
        decoration={markings,mark=at position .55 with {\arrow[draw=black]{>}}}},
    scalarbar/.style={dashed,draw=black, postaction={decorate},
        decoration={markings,mark=at position .55 with {\arrow[draw=black]{<}}}},
    scalarnoarrow/.style={dashed,draw=black},
    electron/.style={draw=black, postaction={decorate},
        decoration={markings,mark=at position .55 with {\arrow[draw=black]{>}}}},
	bigvector/.style={decorate, decoration={snake,amplitude=4pt}, draw},
}
\tikzset{cross/.style={cross out, draw, 
         minimum size=2*(#1-\pgflinewidth), 
         inner sep=0pt, outer sep=0pt}}

\tikzstyle{block} = [draw, rectangle, 
    minimum height=3em, minimum width=6em]

\newcommand{\agl}[2]{\langle#1 \, #2 \rangle}
\newcommand{\sqr}[2]{\lbrack #1 \, #2 \rbrack}

\newcommand{\cO}{\mathcal{O}}
\newcommand{\tb}{\tilde{b}}

\usepackage{float}
\usepackage{marvosym}
\usepackage{empheq}
\usepackage{bbold}

\usepackage{tikz}
\usetikzlibrary{shapes}
\usetikzlibrary{arrows}
\usetikzlibrary{positioning}
\usetikzlibrary{decorations.pathmorphing}
\usetikzlibrary{decorations.pathreplacing}
\usetikzlibrary{decorations.markings}

\begin{document}


\begin{flushright}
	QMUL-PH-19-31\\
	SAGEX-19-29-E\\
\end{flushright}

\vspace{20pt} 

\begin{center}

	{\Large \bf  A note on the  absence of $R^2$ corrections }  \\
	\vspace{0.3 cm} {\Large \bf  to   Newton's potential }

	\vspace{25pt}

	{\mbox {\sf  \!\!\!\!Manuel~Accettulli~Huber, Andreas~Brandhuber, Stefano~De~Angelis and 				Gabriele~Travaglini{\includegraphics[scale=0.05]{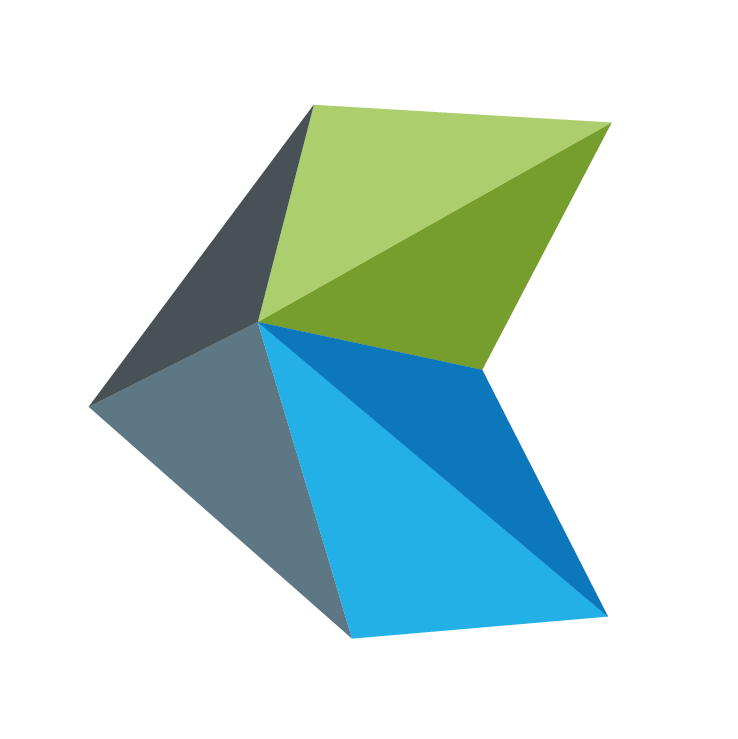}}
	}}
	\vspace{0.5cm}

	\begin{center}
		{\small \em
			Centre for Research in String Theory\\
			School of Physics and Astronomy\\
			Queen Mary University of London\\
			Mile End Road, London E1 4NS, United Kingdom
		}
	\end{center}


	\vspace{40pt}  

	{\bf Abstract}
\end{center}

\vspace{0.3cm}

\noindent

\noindent
We consider Einstein gravity with the addition of $R^2$ and $R^{\mu \nu} R_{\mu \nu}$ interactions  in the context of effective field theory, and the corresponding scattering amplitudes of gravitons and  minimally-coupled heavy scalars.  First, we recover the known fact that graviton amplitudes are the same as in Einstein gravity. Then we show that all amplitudes with two heavy scalars  and an arbitrary number of gravitons are also not affected by these interactions. We prove this by direct computations, using field redefinitions known from earlier applications in string theory, and with a combination of factorisation and power-counting arguments.
Combined with unitarity, these results imply that, in an effective field theory approach,  the  Newtonian potential receives neither classical nor quantum corrections from terms quadratic in the curvature.

\vfill
\hrulefill
\newline
\vspace{-1cm}
${\includegraphics[scale=0.05]{Sagex.jpeg}}$~\!\!{\tt\footnotesize\{m.accettullihuber, a.brandhuber, s.deangelis, g.travaglini\}@qmul.ac.uk}

\setcounter{page}{0}
\thispagestyle{empty}
\newpage


\setcounter{tocdepth}{4}
\hrule height 0.75pt
\tableofcontents
\vspace{0.8cm}
\hrule height 0.75pt
\vspace{1cm}
\setcounter{tocdepth}{2}

\newpage

\section{Introduction}

Much work has been devoted recently to studying the effects of possible modifications of Einstein-Hilbert (EH) gravity, see
 \cite{Clifton:2011jh,Berti:2015itd} for  recent reviews.
Apart from adding a cosmological constant, the  conceptually simplest modifications consist in adding  terms with higher powers of the curvature to the EH action. 
Quadratic and cubic  corrections make an  appearance in the effective gravitational action for closed strings in \cite{Tseytlin:1986zz,Deser:1986xr,Tseytlin:1986ti,Metsaev:1986yb}, and  as 
counterterms at one loop in gravity coupled to matter \cite{tHooft:1974toh}
and at two loops in pure gravity  \cite{Goroff:1985sz,Goroff:1985th}.
At the quadratic level, the independent operators can be taken to be  
$R^2$,  $R^{\mu \nu}R_{\mu \nu}$ and the Gau\ss-Bonnet (GB) combination $R^{\mu \nu \rho \lambda} R_{\mu \nu \rho \lambda} - 4R^{\mu \nu}R_{\mu \nu} + R^2$. The analysis of  \cite{Tseytlin:1986zz,Deser:1986xr,Tseytlin:1986ti} showed that $R^2$ or $R^{\mu \nu}R_{\mu \nu}$  cannot be probed by looking at scattering amplitudes, since they can be removed by  field redefinitions without influencing the S-matrix as a consequence of the S-matrix equivalence theorem, reviewed later, while the GB term, being topological in four dimensions, 
can be discarded.%
\footnote{In $D\!=\!4-2\epsilon$ dimensions, the latter  gives at one loop only finite (quantum) terms which are local and thus do not contribute to the gravitational potential. 
At higher loops the issue should be reconsidered, see \cite{Bern:2015xsa,Bern:2017tuc} for a discussion of the physical (ir)relevance of  evanescent terms. 
}

A related question is whether and how higher-derivative corrections affect the Newtonian potential. The set-up here is that of considering the elastic scattering amplitude of two heavy scalars  minimally coupled to  the gravitational field, from which the form of the gravitational potential can then be extracted \cite{Iwasaki:1971vb,Iwasaki:1971iy,Donoghue:1994dn,BjerrumBohr:2002kt, Neill:2013wsa,Bjerrum-Bohr:2013bxa}. 
The  recent works \cite{Brandhuber:2019qpg,Emond:2019crr} addressed the effect of  terms cubic in the curvature   on the Newtonian potential and particle bending angle,%
\footnote{See  \cite{Cristofoli:2019ewu,Bjerrum-Bohr:2019kec} for an alternative way to extract the two-body Hamiltonian from the scattering amplitude.} 
and in this note we assess the effect of quadratic terms. 
The coupling of the massive scalars to the gravitational field is different than that of the dilaton in string theory, hence the question should be reassessed. Some of these results are probably known but given the renewed interest in the connections between scattering amplitudes and gravitational physics it seems timely to collect these insights also in the light of modern amplitude methods.

We model the two heavy bodies probing the gravitational potential by massive scalars, and the relevant action is 
\begin{equation}
    \label{eq::firstaction}
    S = \int\!\dd[D]x \, \sqrt{-g} \, \Big[ -\frac{2}{\kappa^2} R \, + \, a R^{\mu \nu}R_{\mu \nu} + \, b R^2 \, +\, \frac{1}{2} \sum_{i=1}^{2} \left( \partial_{\mu} \phi_{i} \partial^{\mu} \phi_{i} - m_i^2 \phi_i^2 \right) \Big]\ .
\end{equation}
Note that the scalars are not allowed to propagate in loops, as their sole purpose is to act as massive sources.

One approach, not followed here, is to treat the higher-derivative  corrections exactly, {\it i.e.}~to all orders in $a$ and $b$. 
The analysis  carried out in \cite{Stelle:1976gc,Stelle:1977ry,Julve:1978xn,Tseytlin:1986zz,Metsaev:1986yb,Alvarez-Gaume:2015rwa} shows the presence of additional poles  in the propagator of the linearised metric tensor field $h_{\mu \nu}$:   a massive ghost/tachyon and a massive scalar appear in the spectrum because of the addition of the $R^{\mu \nu} R_{\mu \nu}$ and $R^2$ terms. 
At tree level, this leads to the following corrections to the Newtonian potential   \cite{Stelle:1976gc,Stelle:1977ry}:
\begin{align}
V (r) =-  \frac{\kappa^2}{32 \pi}\frac{M}{r}  \left( 1  - \frac{4}{3} e^{- m_2 r} + \frac{1}{3} e^{- m_0 r}\right) 
\ , 
\end{align}
where 
\begin{align}
m_2 \ = \ \frac{1}{\kappa} \sqrt{-\frac{2}{a}} \ , \qquad m_0 \ = \ \frac{1}{\kappa} \frac{1}{\sqrt {3 b + a}}
\ .
\end{align} 
However, here we take a different route, first advocated in \cite{Donoghue:1994dn},  and treat the Lagrangian \eqref{eq::firstaction} as that of an effective field theory
\cite{Simon:1990ic, Simon:1990jn, Donoghue:1995cz}, 
with the 
 dimensionless parameters $a$ and $b$ considered  as small. In this case, the masses of the new particles alluded to earlier would be   above the cut-off of our theory, hence these particles should not be included as genuinely propagating states.%
 \footnote{ Even for $a$ and $b$ being of $\mathcal{O}(1)$,
these masses would be of the order of the Planck mass, where a whole tower of higher-derivative terms would have to be included.}  
The effect of these terms in our treatment will be that of introducing new vertices,  including two-point vertices, which give rise to local interaction terms, with the spectrum being unmodified compared to that of EH gravity. 
In particular, in the effective field theory approach the new Yukawa potentials induced by the quadratic terms are absent at tree level \cite{Donoghue:1995cz}. This is best seen  in momentum space, where the massive propagators are replaced by a polynomial in the momentum transfer squared, which in turn leads to local terms which give no contribution to long-range physics.  

Importantly, this is not the end of the story since further classical (and of course quantum) corrections can emerge from loop diagrams \cite{Holstein:2004dn}.
A great simplification stems from the fact that we are interested only in effects on low-energy physics -- the 
classical and quantum corrections to the potential. These can  only arise from terms in the scattering amplitude that are non-analytic in the momentum transfer squared $q^2:=(p_1+p_2)^2$ between the two massive scalars  \cite{Donoghue:1994dn, Holstein:2004dn}  and can be efficiently captured using unitarity-based methods \cite{Bern:1994cg,Bern:1994zx}. The latter  approach was used efficiently in 
 \cite{Neill:2013wsa,Bjerrum-Bohr:2013bxa} and \cite{Bjerrum-Bohr:2014zsa, Bjerrum-Bohr:2016hpa, Bai:2016ivl, Chi:2019owc} to extract the classical and quantum corrections at  $\mathcal{O}(G_N^2)$ to the Newton potential and particle bending angle, respectively, where $G_N:=\kappa^2/ ( 32 \pi)$ is Newton's constant. We also note the recent works \cite{Bern:2019nnu,Bern:2019crd}, where the conservative hamiltonian for binary systems was extracted at    $\mathcal{O} ( G_N^3)$ from two-loop amplitude computations. 
 Therefore we only need to focus on unitarity cuts in this channel.
 
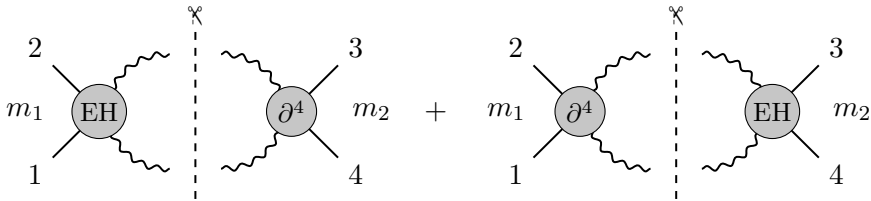
\begin{figure}[h!]
    \centering
    \begin{tikzpicture}[scale=12]

        \def\x{0};
        \def\y{0};
        \node at (0+\x,0+\y) (left) {};
        \node at (-2pt+\x,2pt+\y) (oneA) {2};
        \node at (-2pt+\x,-2pt+\y) (oneB) {1};
        \node at (1.7pt+\x,1.5pt+\y) (top){};
        \node at (1.7pt+\x,-1.5pt+\y) (bottom){};

        \draw [line width=.75,vector] (2.2pt,1.8pt) arc (90:180:1.8pt);
        \draw [line width=.75,vector] (2.2pt,-1.8pt) arc (270:180:1.8pt);
        \draw [line width=.75] (oneA) -- (left);
        \draw [line width=.75] (oneB) -- (left);
        \node at (0+\x,0+\y) [draw, fill=gray!45!white, circle, inner sep=1.7pt]{\small EH};

        \draw [dashed, line width=.75] (3pt+\x,2.5pt+\y) -- (3pt+\x,-3pt);
        \node at (3.1pt+\x,2.7pt+\y)[rotate around={-90:(1pt,2.5pt)}]{\Cutright};
        \node at (-2.3pt+\x,0pt+\y) {$m_1$};

        \def\x{6pt};
        \def\y{0};

        \node at (0+\x,0+\y) (left) {};
        \node at (2pt+\x,2pt+\y) (oneA) {3};
        \node at (2pt+\x,-2pt+\y) (oneB) {4};
        \node at (-1.7pt+\x,1.5pt+\y) (top){};
        \node at (-1.7pt+\x,-1.5pt+\y) (bottom){};

        \draw [line width=.75,vector] (-2.2pt+\x,1.8pt) arc (90:0:1.8pt);
        \draw [line width=.75,vector] (-2.2pt+\x,-1.8pt) arc (270:360:1.8pt);
        \draw [line width=.75] (oneA) -- (left);
        \draw [line width=.75] (oneB) -- (left);
        \node at (0+\x,0+\y) [draw, fill=gray!45!white, circle, inner sep=2pt]{\small $\partial^4$};
        \node at (2.5pt+\x,0+\y) {$m_2$};

        \node at (4.5pt+\x,0+\y){+};

    \def\x{15pt};
        \def\y{0};
        \node at (0+\x,0+\y) (left) {};
        \node at (-2pt+\x,2pt+\y) (oneA) {2};
        \node at (-2pt+\x,-2pt+\y) (oneB) {1};
        \node at (1.7pt+\x,1.5pt+\y) (top){};
        \node at (1.7pt+\x,-1.5pt+\y) (bottom){};

        \draw [line width=.75,vector] (2.2pt+\x,1.8pt+\y) arc (90:180:1.8pt);
        \draw [line width=.75,vector] (2.2pt+\x,-1.8pt+\y) arc (270:180:1.8pt);
        \draw [line width=.75] (oneA) -- (left);
        \draw [line width=.75] (oneB) -- (left);
        \node at (0+\x,0+\y) [draw, fill=gray!45!white, circle, inner sep=2pt]{\small $\partial^4$};

        \draw [dashed, line width=.75] (3pt+\x,2.5pt+\y) -- (3pt+\x,-3pt);
        \node at (3.1pt+\x,2.7pt+\y)[rotate around={-90:(1pt,2.5pt)}]{\Cutright};
        \node at (-2.3pt+\x,0pt+\y) {$m_1$};

        \def\x{21pt};
        \def\y{0};

        \node at (0+\x,0+\y) (left) {};
        \node at (2pt+\x,2pt+\y) (oneA) {3};
        \node at (2pt+\x,-2pt+\y) (oneB) {4};
        \node at (-1.7pt+\x,1.5pt+\y) (top){};
        \node at (-1.7pt+\x,-1.5pt+\y) (bottom){};

        \draw [line width=.75,vector] (-2.2pt+\x,1.8pt+\y) arc (90:0:1.8pt);
        \draw [line width=.75,vector] (-2.2pt+\x,-1.8pt+\y) arc (270:360:1.8pt);
        \draw [line width=.75] (oneA) -- (left);
        \draw [line width=.75] (oneB) -- (left);
        \node at (0+\x,0+\y) [draw, fill=gray!45!white, circle, inner sep=1.7pt]{\small EH};
        \node at (2.5pt+\x,0+\y) {$m_2$};
    \end{tikzpicture}
     \caption{The one-loop unitarity cut in the $q^2$-channel contributing to the massive scalar scattering. Here the $\partial^4$ blob denotes the amplitude with one insertion of either $R^2$ or $R^{\mu \nu}R_{\mu \nu}$. 
     }
\label{cuts}
\end{figure}

At one loop, we  have to consider a two-particle cut such  as that in Figure \ref{cuts}. As the figure shows, at this loop order there are two  building  blocks:   the tree-level two-scalar/two-graviton amplitudes in  EH, and the same amplitudes with one insertion of the quadratic corrections to the action, see Figure~\ref{cuts}.

There are only two Feynman diagrams contributing to the latter and it turns out that their sum is zero both for $R^{\mu \nu} R_{\mu \nu}$ and $R^2$, see Figure~\ref{pallo}.  This result implies that  the contribution to the scalar potential  to first order in $a$ or $b$ is also zero at 2PM, since the
amplitude with one insertion of the quadratic corrections vanishes.  
\begin{figure}[h!]
    \centering
    \begin{tikzpicture}[scale=12]

        \def\x{-7pt};
        \def\y{0};

        \node at (0+\x,0+\y) (left) {};
        \node at (2pt+\x,2pt+\y) (oneA) {};
        \node at (2pt+\x,-2pt+\y) (oneB) {};
        \node at (-1.7pt+\x,1.5pt+\y) (top){};
        \node at (-1.7pt+\x,-1.5pt+\y) (bottom){};

        \draw [line width=.75,vector] (-2.2pt+\x,1.8pt+\y) arc (90:0:1.8pt);
        \draw [line width=.75,vector] (-2.2pt+\x,-1.8pt+\y) arc (270:360:1.8pt);
        \draw [line width=.75] (oneA) -- (left);
        \draw [line width=.75] (oneB) -- (left);
        \node at (0+\x,0+\y) [draw, fill=gray!45!white, circle, inner sep=2pt]{\small $\partial^4$};

        \node at (3pt+\x,0+\y){=};

        \def\x{-0.5pt};
        \def\y{0};

        \draw [thick](-1.3pt+\x,2.2pt+\y) -- (-1.7pt+\x,2.2pt+\y) -- (-1.7pt+\x,-2.2pt+\y) -- (-1.3pt+\x,-2.2pt+\y);

        \def\x{0};
        \node at (0+\x,0+\y) (centro) {};
        \node at (2pt+\x,0+\y) (tre) {};
        \node at (2pt+\x,-0.6pt+\y) {$\mu \nu$};
        \node at (-1pt+\x,2pt+\y) (due) {};
        \node at (-1pt+\x,-2pt+\y) (uno) {};

        \draw [vector, line width=.75] (uno) -- (centro.center);
        \draw [vector, line width=.75] (due) -- (centro.center);
        \draw [vector, line width=.75] (centro.center) -- (tre);
        \node at (centro) [draw,circle, fill=black, inner sep=1.5pt] {};

        \node at (3.5pt,0) {+};

        \def\x{6pt};
        \def\y{0};
        \node at (0+\x,0+\y) (centro) {};
        \node at (3pt+\x,0+\y) (tre) {};
        \node at (3pt+\x,-0.6pt+\y) {$\mu \nu$};
        \node at (-1pt+\x,2pt+\y) (due) {};
        \node at (-1pt+\x,-2pt+\y) (uno) {};

        \draw [vector, line width=.75] (uno) -- (centro.center);
        \draw [vector, line width=.75] (due) -- (centro.center);
        \draw [vector, line width=.75] (centro.center) -- (tre);
        \node at (1.5pt+\x,0+\y) [draw,circle, fill=black, inner sep=1.5pt] {};

        \def\x{8.5pt}
        \draw [thick](1.3pt+\x,2.2pt+\y) -- (1.7pt+\x,2.2pt+\y) -- (1.7pt+\x,-2.2pt+\y) -- (1.3pt+\x,-2.2pt+\y);

        \def\x{13pt};
        \def\y{0};

        \node at (0+\x,0+\y) (centro) {};
        \node at (-2pt+\x,0+\y) (tre) {};
        \node at (-2pt+\x,0.6pt+\y) {$\mu \nu$};
        \node at (1pt+\x,2pt+\y) (due) {};
        \node at (1pt+\x,-2pt+\y) (uno) {};

        \draw [line width=.75] (uno) -- (centro.center);
        \draw [line width=.75] (due) -- (centro.center);
        \draw [vector, line width=.75] (centro.center) -- (tre);

        \draw [thick,decorate,decoration={brace,amplitude=10pt,mirror,raise=4pt},yshift=0pt] (-2.2pt,-2.3pt) -- (10.2pt,-2.3pt);

        \node at (4pt,-4pt) {\large $=0$};

    \end{tikzpicture}
    \caption{The sum of the two diagrams contributing to the two-scalar two-graviton amplitudes to  first order in $R^2$. The same result holds for $R^{\mu \nu} R_{\mu \nu}$. All external on-shell states are in $D$ dimensions, the legs labelled by Lorentz indices are off-shell. The relevant Feynman rules can be found in Appendix~\ref{sec::Feynmanrules}.}
    \label{pallo}
\end{figure}

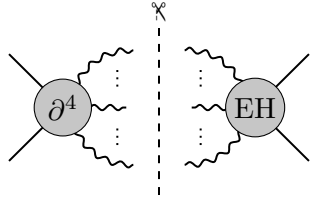
\begin{figure}[h]
\centering
\begin{tikzpicture}[scale=12]
        \def\x{0};
        \def\y{0};
        \node at (0+\x,0+\y) (left) {};
        \node at (-2pt+\x,2pt+\y) (oneA) {};
        \node at (-2pt+\x,-2pt+\y) (oneB) {};
        \node at (2.3pt+\x,0+\y) (three) {};
        \node at (1.7pt+\x,1.5pt+\y) (top){};
        \node at (1.7pt+\x,-1.5pt+\y) (bottom){};

        \draw [line width=.75,vector] (2.2pt,1.8pt) arc (90:180:1.8pt);
        \draw [line width=.75,vector] (2.2pt,-1.8pt) arc (270:180:1.8pt);
        \draw [line width=.75] (oneA) -- (left);
        \draw [line width=.75] (oneB) -- (left);
        \draw [dotted, line width=.75] (top) -- + (0,-1pt);
        \draw [dotted, line width=.75] (bottom) -- + (0,1pt);
        \draw [vector,line width=.75] (left) -- (three);
        \node at (0+\x,0+\y) [draw, fill=gray!45!white, circle, inner sep=2.5pt]{$\partial^4$};

        \draw [dashed, line width=.75] (3pt+\x,2.5pt+\y) -- (3pt+\x,-3pt);
        \node at (3.1pt+\x,2.7pt+\y)[rotate around={-90:(1pt,2.5pt)}]{\Cutright};

        \def\x{6pt};
        \def\y{0};

        \node at (0+\x,0+\y) (left) {};
        \node at (2pt+\x,2pt+\y) (oneA) {};
        \node at (2pt+\x,-2pt+\y) (oneB) {};
        \node at (-2.3pt+\x,0+\y) (three) {};
        \node at (-1.7pt+\x,1.5pt+\y) (top){};
        \node at (-1.7pt+\x,-1.5pt+\y) (bottom){};

        \draw [line width=.75,vector] (-2.2pt+\x,1.8pt) arc (90:0:1.8pt);
        \draw [line width=.75,vector] (-2.2pt+\x,-1.8pt) arc (270:360:1.8pt);
        \draw [line width=.75] (oneA) -- (left);
        \draw [line width=.75] (oneB) -- (left);
        \draw [dotted, line width=.75] (top) -- + (0,-1pt);
        \draw [dotted, line width=.75] (bottom) -- + (0,1pt);
        \draw [vector,line width=.75] (left) -- (three);
        \node at (0+\x,0+\y)  [draw, fill=gray!45!white, circle, inner sep=1.8pt]{EH};
    \end{tikzpicture}
\caption{An example of higher-loop cut diagram contributing to the Newtonian potential. 
The $\partial^4$ symbol denotes an insertion of either $R^2$ or $R^{\mu \nu}R_{\mu \nu}$. }
    \label{pallino}
   \end{figure}

 This  argument can be extended straightforwardly to higher loops. In practice, the focus will be on cut diagrams such as the one depicted in Figure~\ref{pallino}. One of the amplitudes in the cut is in the background of a quadratic correction in the curvature, while the other is a standard EH amplitude. The types of amplitudes that will be needed are: amplitudes with two scalars and an arbitrary number of gravitons, and amplitudes only made of gravitons.%
 \footnote{See Figure 1 of \cite{Bern:2019nnu} for sample two-loop cut diagrams.}

The result of this note is that such amplitudes in a four-derivative background are all zero, hence do not affect Newton's potential to any order in $G_N$. 
We prove this in three  ways: 
\begin{itemize}
\item[{\bf 1.}] Using a redefinition of the metric, in conjunction with the S-matrix equivalence theorem, similarly to what was done in \cite{Tseytlin:1986zz,Deser:1986xr,Tseytlin:1986ti}. Results obtained here are valid up to linear order in the small parameters $a$, and $b$, which is perfectly sufficient from an effective field theory point of view. This argument is valid in $D$ dimensions, hence it  lends itself to  an application of $D$-dimensional unitarity; 
\item[{\bf 2.}] Using a combination of dimensional analysis and little-group scaling. This argument is valid in four dimensions, as far as the gravitons are concerned; 
\item[{\bf 3.}] Using a diagrammatic argument, which turns out to be valid for any number of insertions of the higher-derivative couplings. 
\end{itemize}
These three approaches will be discussed in turn in the three following sections. We also include an appendix, containing  the  Feynman rules needed in the calculations.

\section{Taming quadratic   terms with field redefinitions}\label{sec:fieldredefinition}

In this section we show that $n$-point gravitons and two-scalar/$n$-graviton amplitudes in the background of a term quadratic in the curvature are zero. 
A key ingredient in our proof is the S-matrix equivalence theorem.%
\footnote{This beautiful theorem has a long-winded history that we will not attempt to retrace here. 
An incomplete list of relevant works include \cite{Borchers:1962cbs,Coleman:1967vs,Coleman:1969sm,tHooft:1974toh, Bergere:1975tr,Haag:1992hx,Ettle:2006bw,Brandhuber:2006bf}.
}
According to this theorem, fairly generic field redefinitions do not alter the S-matrix. In the context of the effective  action of   string theory, this has been used   
to show that  terms quadratic   in $R$ or $R_{\mu \nu}$  \cite{Tseytlin:1986zz,Deser:1986xr,Tseytlin:1986ti} or 
containing any power of $R$ or $R_{\mu \nu}$ (apart from the EH term) 
\cite{Metsaev:1986yb} do not affect the S-matrix -- they can be redefined away. 
With a similar logic, we introduce the following 
local field redefinition  of the metric: 
\begin{equation}
    g_{\mu \nu} \rightarrow g_{\mu \nu} + \alpha_{1} g_{\mu \nu} R+\alpha_{2} R_{\mu \nu}+\sum_{i=1}^{2} \beta_{1}^{(i)} \partial_{\mu}{\phi_{i}} \partial_{\nu}{\phi_{i}}+\beta_{2}^{(i)} g_{\mu \nu} \partial_{\sigma}{\phi_{i}} \partial^{\sigma}{\phi_{i}}+\beta_{3}^{(i)} g_{\mu \nu} \phi_{i}^2\, . 
\end{equation}
The main point here  is that we can fix the  $\alpha$ parameters  by requiring the vanishing of the coefficients of the $R^2$ and $R^{\mu \nu} R_{\mu \nu}$  interactions, while  the $\beta$ parameters  can be fixed in such a way that no  non-minimal coupling between gravity and the scalar fields are  generated. To first order in the parameters $a$ and $b$ these non-minimal interactions have the form  $R \partial_{\mu}{\phi_{i}} \partial^{\mu}{\phi_{i}}$, $R^{\mu \nu} \partial_{\mu}{\phi_{i}} \partial_{\nu}{\phi_{i}}$ and $R \phi_{i}^{2}$. To first order in $a$ and $b$, the solution is: 
\begin{equation}
    \alpha_{1} = \frac{ a + 2 b }{2 (D-2)} \kappa^2\ , \hspace{1cm} \alpha_{2} = -\frac{a \kappa^2}{2} \ ,
\end{equation}
and
\begin{equation}
    \beta_{1}^{(i)} = -\frac{a \kappa^4}{8}\ , \hspace{1cm} \beta_{2}^{(i)} = \frac{ a + 2 b }{8 (D-2)} \kappa^4\ , \hspace{1cm} \beta_{3}^{(i)}=- \frac{a + D b}{2 (D-2)^2} m_{i}^2 \kappa^4\ .
\end{equation}
Under the  field redefinitions specified above, the original action \eqref{eq::firstaction} becomes
\begin{equation}
\label{newac}
    \begin{split}
        S'& =  \int\!\dd[D]x \, \sqrt{-g} \, \bigg[ -\frac{2}{\kappa^2} R +\, \frac{1}{2} \sum_{i=1}^{2} \left( \partial_{\mu} \phi_{i} \partial^{\mu} \phi_{i} - m_i^2 \phi_i^2 \right) + {\kappa}^{4}\, \frac{D (a + D b)}{4(D-2)^2} \Big( \sum_{i=1}^{2} m_{i}^{2} \phi_{i}^{2}
        \Big)^2\\[.5em]
        &- {\kappa}^{4}\, \frac{a + D b}{4 (D-2)} \Big( \sum_{i=1}^{2} m_{i}^{2} \phi_{i}^{2}
        \Big) \Big( \sum_{i=1}^{2} \partial_{\mu}{\phi_{i}} \partial^{\mu}{\phi_{i}} 
        \Big)+ {\kappa}^{4}\, \frac{a+b}{16} \Big( \sum_{i=1}^{2} \partial_{\mu}{\phi_{i}} \partial^{\mu}{\phi_{i}}
        \Big)^2 + \mathcal{O}(a^2,b^2, a b )\bigg]\ . 
    \end{split}
\end{equation}
By the equivalence theorem, $S$ and $S^\prime$ lead to the same S-matrix. 
From the  new  action \eqref{newac}  it is now manifest  that no corrections to the EH (two-scalar) $n$-graviton amplitudes are  generated.

It is also interesting to note that 
the field redefinition introduces contact terms for the four-scalar amplitude. One can easily check that the result for this quantity from \eqref{eq::firstaction} matches exactly the new four-point vertex:
\begin{equation}
    \frac{\delta^4 S'}{\delta\phi_1 \delta \phi_1 \delta\phi_2\delta\phi_2} =\hspace{0.3cm}
    \begin{tikzpicture}[scale=12,baseline={([yshift=-1mm]center.base)}]
        \clip (-2.5pt,-2.5pt) rectangle (2.5pt,2.5pt);
        \node at (-2pt,-2pt) (one){$\phi_1$};
        \node at (-2pt,2pt) (two) {$\phi_1$};
        \node at (2pt,2pt) (three) {$\phi_2$};
        \node at (2pt,-2pt) (four) {$\phi_2$};
        \draw[line width=.75] (one) -- (three);
        \draw[line width=.75] (two) -- (four);
        \node at (0,0)[circle,draw,fill=gray!45!white, inner sep=7pt] (center) {};
    \end{tikzpicture}
    \hspace{0.2cm} = \hspace{0.2cm}
    \begin{tikzpicture}[scale=12,baseline={([yshift=-1mm]center.base)}]
        \def\x{0};
        \def\y{0};
        \node at (-1pt+\x,0+\y) (one) {};
        \node at (1pt+\x,0+\y) (two) {};
        \node at (-2.5pt+\x,2pt+\y) (oneA) {$\phi_1$};
        \node at (-2.5pt+\x,-2pt+\y) (oneB) {$\phi_1$};
        \node at (2.5pt+\x,2pt+\y) (twoA) {$\phi_2$};
        \node at (2.5pt+\x,-2pt+\y) (twoB) {$\phi_2$};

        \draw[vector, line width=.75] (one.center) -- (two.center);
        \node at (0+\x,0+\y) [draw, circle,fill=black,inner sep=1.5pt]{};
        \draw [line width=.75] (oneA) -- (one.center);
        \draw [line width=.75] (oneB) -- (one.center);
        \draw [line width=.75] (twoA) -- (two.center);
        \draw [line width=.75] (twoB) -- (two.center);

    \end{tikzpicture}
    \hspace{0.2cm} + \hspace{0.2cm}
    \begin{tikzpicture}[scale=12,baseline={([yshift=-1mm]Center.base)}]
        \def\x{0};
        \def\y{0};
        \node at (-1pt+\x,0+\y) (one) {};
        \node at (1pt+\x,0+\y) (two) {};
        \node at (-2.5pt+\x,2pt+\y) (oneA) {$\phi_1$};
        \node at (-2.5pt+\x,-2pt+\y) (oneB) {$\phi_1$};
        \node at (2.5pt+\x,2pt+\y) (twoA) {$\phi_2$};
        \node at (2.5pt+\x,-2pt+\y) (twoB) {$\phi_2$};
        \node at (0+\x,0+\y) (Center){};

        \draw[vector, line width=.75] (one.center) -- (Center.center);
        \draw[vector, line width=.75] (two.center) -- (Center.center);
        \node at (0+\x,0+\y) [draw, cross,inner sep=3pt,line width=1.75]{};
        \draw [line width=.75] (oneA) -- (one.center);
        \draw [line width=.75] (oneB) -- (one.center);
        \draw [line width=.75] (twoA) -- (two.center);
        \draw [line width=.75] (twoB) -- (two.center);

    \end{tikzpicture}
\end{equation}
where the dot and the cross denote the insertion of $R^2$ and $R^{\mu\nu} R_{\mu \nu}$, respectively.

We truncated the action $S'$ to linear order in $a$ and $b$. 
Keeping higher orders in these parameters would imply that  higher-derivative terms   such as  $R\, \Box R$, $R^3$ and $R R^{\mu \nu} R_{\mu \nu}$ appear in the new action. These  could in turn be eliminated by adding further  terms (involving more derivatives) to the field redefinition.
Note that the  above-mentioned contact terms of the scalar fields (which do not affect the computation of corrections to the Newton potential), and contractions of three or more Riemann tensors (which lead to genuine modifications of the Newtonian potential \cite{Brandhuber:2019qpg,Emond:2019crr}), cannot be eliminated in this way.
Finally, we note that  the field-redefinition argument we discussed can in principle be applied to a wider class of  terms including non-minimal couplings of the scalars to gravity.

\section{Taming quadratic terms with  amplitude techniques}
\label{section:3}

In this section we address the question of the absence  of two-scalar/$n$-graviton and $n$-graviton amplitudes induced by terms quadratic in $R$ or $R_{\mu \nu}$ from a modern amplitude perspective. 
This viewpoint  allows to address this question  to higher orders in the four-derivative couplings, and furthermore has the  advantage 
of treating all four-derivative interactions  in \eqref{eq::firstaction} in one go. 
 For this reason, in this section we will refer in general to any of the four-derivative interactions as $R^2$ and to any of the 
two associated couplings as $\tilde{b}$.

The argument is two-fold. First, we show the absence of possible factorisations for an amplitude with two scalars and two gravitons. 
Next, we show that no two-scalar/$n$-graviton contact terms, unseen by factorisation,  are present. Together, these imply the absence of two-scalar/$n$-graviton amplitudes with one insertion of $R^2$. 
We address these two parts in turn.

\subsubsection*{Absence of factorisation channels}

A two-scalar/two-graviton amplitude in an $R^2$ background could factorise onto an EH  three-point scalar-scalar-graviton amplitude and a three-graviton amplitude produced by an $R^2$ interaction. However it is elementary to show that in four dimensions
$R^2$ couplings cannot modify the three-graviton amplitude, and all three-point amplitudes arise from either EH gravity or  a six-derivative modification  involving $R^{\alpha \beta}\,_{\mu \nu} R^{\mu \nu}\,_{\rho \sigma} R^{\rho \sigma}\,_{\alpha \beta}$.

Little-group scaling, combined with considerations of the mass-dimension of the couplings, constrains the most general form of the three-graviton amplitudes. We begin by considering  the three-graviton amplitude $A_3(1^{++},2^{++},3^{--})$. It is well known that there are  only two possible helicity structures for this amplitude,  
\begin{equation}
    A_3(1^{++},2^{++},3^{--}) \, \sim \,  \frac{\sqr{1}{2}^6}{\sqr{2}{3}^2\sqr{3}{1}^2} \>
 \quad \text{or}\quad  
    \widetilde{A}_3(1^{++},2^{++},3^{--}) \, \sim  \,  \frac{\agl{2}{3}^2\agl{3}{1}^2}{\agl{1}{2}^6} \> . 
\end{equation}
Purely on dimensional grounds, the first amplitude arises from a two-derivative interaction, such as EH gravity, while the latter would require a non-local interaction in the theory and should be discarded. 
Next consider the all-plus helicity configuration $A_3(1^{++},2^{++},3^{++})$. Here one has two possibilities, 
\begin{equation}
    A_3(1^{++},2^{++},3^{++})\, \sim \, \sqr{1}{2}^2 \sqr{2}{3}^2 \sqr{3}{1}^2 \> 
\quad \text{or}
\quad 
  \widetilde{A}_3(1^{++},2^{++},3^{++})\, \sim \,  \frac{1}{\agl{1}{2}^2 \agl{2}{3}^2 \agl{3}{1}^2} \> .
\end{equation}
We can immediately discard the second one  from the request of locality.
As for the first, it arises from six-derivative interaction terms such as   $R^{\alpha \beta}\,_{\mu \nu} R^{\mu \nu}\,_{\rho \sigma} R^{\rho \sigma}\,_{\alpha \beta}$ or $R^{\alpha}\,_{\mu}\,^{\beta}\,_{\nu} R^{\mu}\,_{\rho}\,^{\nu}\,_{\sigma} R^{\rho}\,_{\alpha}\,^{\sigma}\,_{\beta}$, but not from those terms which can be eliminated by a field redefinition, {\it e.g.} $R\, \square R$ or $R R^{\mu \nu} R_{\mu \nu}$.
As a consequence, the addition of a four-derivative interactions  to the action with mass-dimension zero coupling $\tb$ cannot generate  a three-point amplitude in four dimensions.
Hence, any two-scalar/two-graviton amplitude in the $R^2$ background can only be a contact term.  

\subsubsection*{Absence of contact terms}

We still have the possibility of a two-scalar/two-graviton contact term, which via factorisation would give rise to non-trivial higher-point amplitudes. In the following we will  show that such  contact terms are absent for amplitudes with any number of gravitons and up to two scalars. 

First we consider pure graviton amplitudes at tree level, starting from four external particles. We showed that the three-graviton amplitude is unaffected by any $R^2$ insertion, hence  no  factorisation channel for the four-point amplitude is available. The remaining task is to exclude potential  contact terms. The latter can in general be written schematically at a given multiplicity $n$ and order $B$ in $\tb$ as
\begin{align}
\mathcal{M}_{\rm contact} \ \propto  \ \tb^B \kappa^{ n-2 + 2B} \prod_{i=1}^n \lambda^{\otimes a_i}_{i} \tilde\lambda^{\otimes s_i}_{i}
\ .
\end{align}
The $\kappa^{2B}$ comes from the additional powers in $\kappa$ carried by each of the $R^2$ insertions  (with respect to EH) 
along with a single power of $\tb$.%
\footnote{Recall that  we are interested in the effect of the four-derivative interactions on the amplitude, hence  $B>0$.}
Dimensional analysis implies that
\begin{equation}
c+ \frac{1}{2} \sum_{i=1}^n (a_i + s_i)  = \ 4-n\ , 
\end{equation}
where $c = 2 - n - 2B$ is the overall dimension of the couplings.
This can be rewritten as
\begin{equation}\label{eq:constraint}
    \sum_{i=1}^n (a_i + s_i) = 4 + 4 B\ .
\end{equation}
A further constraint comes from little-group scaling, which requires
\begin{equation}\label{eq:hel}
    - a_i + s_i  = 2 h_i \> ,
\end{equation}
where $h_i$ is the helicity of particle $i$. From~\eqref{eq:hel} we also have the constraint that either $a_{i} \geq 4$ or $s_{i} \geq 4$, corresponding  to the helicity of  graviton $i$  being  minus or plus two, respectively.

Specialising now to $n=4$ we see that the latter constraint cannot be satisfied along with~\eqref{eq:constraint} for $B<3$, thus it is not possible to build contact terms with $B=1,2$. This means that up to second order in $\tb$ it is impossible to write down any contact term contribution  to the four-graviton amplitude coming from an $R^2$ interaction.
At $B=3$ the argument breaks down because three insertions of $R^2$ terms can be mimicked by one insertion of the Riemann tensor to the fourth power, which gives rise to a non-vanishing amplitude.

The final step is to  recursively extend the argument to $n>4$. 
We only need to exclude a contact term. It is immediate to realise that 
the combined constraints~\eqref{eq:constraint} and~\eqref{eq:hel} cannot be satisfied for $B<n-1$, hence contact terms up to order $n-2$ in $\tb$ are ruled out. Since the recursive argument starts from three- and four-graviton amplitudes, we conclude that all $n$-graviton amplitudes are unaffected by $R^2$ contributions up to order $2$ in $\tb$. 

Dimensional analysis was sufficient to show that up to $\cO(\tb^2)$ there are no $R^2$ corrections to $n$-graviton amplitudes. In order to push our considerations to even higher orders  in $\tb^2$ we need to invoke additional diagrammatic arguments, which will be detailed in the next section.

The discussion presented so far can   be  generalised with ease to  two-scalar/$n$-graviton amplitudes. 
Once again the three-point minimal interaction involving a single graviton and two scalars is unaffected by the $R^2$ couplings, which rules out factorisation channels in the four-point amplitude. 
The general form of the contact term now becomes
\begin{align}
\mathcal{M}_{\rm contact} \ \sim \ \tb^B \kappa^{ n-2 + 2B}\ {\rm M}(q_{i})\ \prod_{i=3}^{n} \lambda^{\otimes a_i}_{i} \tilde\lambda^{\otimes s_i}_{i}
\ , 
\end{align}
where ${\rm M}$ is a quadratic combination of the momenta $q_{i}$ of the scalars, labelled by $i=1,2$,   arising from their minimal  coupling to gravity. Then $[{\rm M}] = 2$, and repeating the same dimensional analysis  as before  we find
\begin{equation}
\label{ciccio}
    \sum_{i=1}^{n-2} (a_i + s_i) = 4 B\ .
\end{equation}
For $B=1$ the constraints \eqref{ciccio} and \eqref{eq:hel} cannot be satisfied simultaneously, for any number of gravitons. As a side remark, note that this argument does not prevent the appearance of amplitudes with four scalars,  because these amplitudes require a double insertion of ${\rm M}(q_{i})$ with $i=1,\ldots , 4$. This is also  in complete agreement with  our approach based on  field redefinitions -- indeed \eqref{newac} does generate a four-scalar amplitude. 

{ In summary, we have shown the absence of $R^2$ corrections at linear order in the coupling $\tilde{b}$ for all tree amplitudes with $n$ gravitons and up to two scalars. As argued in 
the Introduction, 
 this implies via unitarity the absence of all $R^2$ corrections to the Newtonian potential, both at the classical and quantum level.}

\section{Beyond linear order in $a$~and~$b$ via diagrammatics}

As discussed earlier, from an on-shell perspective it is not possible to push our general considerations further than the first and second order in the $R^2$ and $R^{\mu \nu}R_{
\mu \nu}$ couplings for two-scalar/$n$-graviton or $n$-graviton amplitudes, respectively (again collectively denoted as $\partial^4$ in the figures below). In this section we show how, combining on-shell arguments with diagrammatic insights, we can refine our earlier discussion to include  higher-order corrections in $a$ and $b$.  We consider first the two-scalar two-graviton amplitude at order $\tb$:

\begin{equation}\label{eq:22pointB}
    \begin{tikzpicture}[scale=12,baseline={([yshift=-1mm]center.base)}]
        \clip (-3pt,-3pt) rectangle (3.3pt,3pt);
        \node at (0,0)(center){};
        \node at (-2pt,-2pt) (s1){};
        \node at (-2pt,2pt)(s2){};
        \node at (2pt,2pt) (g1){};
        \node at (2pt,-2pt) (g2){};
        \node at (-2.4pt,-1.8pt) {$1$};
        \node at (-2.4pt,1.8pt){$2$};
        \node at (2.7pt,1.8pt) {$3^{h_3}$};
        \node at (2.7pt,-1.8pt) {$4^{h_4}$};

        \draw [decorate, decoration={snake, amplitude=1pt, segment length=6pt},line width=.75] (g1) -- (center);
        \draw [decorate, decoration={snake, amplitude=1pt, segment length=6pt},line width=.75]  (g2) -- (center);
        \draw [line width=.75]  (s1) -- (center);
        \draw [line width=.75]  (s2) -- (center);

        \node at (0,0) [circle, draw, fill=gray!45!white, inner sep=2pt]{\small $\partial^4$};
    \end{tikzpicture}
    \hspace{0.3cm}\overset{(\tb)}{=}\> \sum_{\rm perm} \> \left(
    \frac{1}{4}
    \begin{tikzpicture}[scale=12,baseline={([yshift=-1mm]center.base)}]
    \clip (-3pt,-2pt) rectangle (3pt,2pt);
    \def\x{0};
        \def\y{0};
        \node at (-1pt+\x,0+\y) (one) {};
        \node at (1pt+\x,0+\y) (two) {};
        \node at (-2.5pt+\x,2pt+\y) (oneA) {};
        \node at (-2.5pt+\x,-2pt+\y) (oneB) {};
        \node at (2.5pt+\x,2pt+\y) (twoA) {};
        \node at (2.5pt+\x,-2pt+\y) (twoB) {};

        \draw [decorate, decoration={snake, amplitude=1pt, segment length=6pt},line width=.75]  (one.center) -- (two.center);
        \draw [line width=.75]  (oneA) -- (one.center);
        \draw [line width=.75]  (oneB) -- (one.center);
        \draw [decorate, decoration={snake, amplitude=1pt, segment length=6pt},line width=.75]  (twoA) -- (two.center);
        \draw [decorate, decoration={snake, amplitude=1pt, segment length=6pt},line width=.75]  (twoB) -- (two.center);
        \node at (1pt+\x,0+\y)[draw, circle,fill=black,inner sep=1.5pt] {};
    \end{tikzpicture}
    +\frac{1}{4}
     \begin{tikzpicture}[scale=12,baseline={([yshift=-1mm]center.base)}]
    \clip (-3pt,-2pt) rectangle (3pt,2pt);
    \def\x{0};
        \def\y{0};
        \node at (-1pt+\x,0+\y) (one) {};
        \node at (1pt+\x,0+\y) (two) {};
        \node at (-2.5pt+\x,2pt+\y) (oneA) {};
        \node at (-2.5pt+\x,-2pt+\y) (oneB) {};
        \node at (2.5pt+\x,2pt+\y) (twoA) {};
        \node at (2.5pt+\x,-2pt+\y) (twoB) {};

        \draw [decorate, decoration={snake, amplitude=1pt, segment length=6pt},line width=.75]  (one.center) -- (two.center);
        \draw [line width=.75]  (oneA) -- (one.center);
        \draw [line width=.75]  (oneB) -- (one.center);
        \draw [decorate, decoration={snake, amplitude=1pt, segment length=6pt},line width=.75]  (twoA) -- (two.center);
        \draw [decorate, decoration={snake, amplitude=1pt, segment length=6pt},line width=.75]  (twoB) -- (two.center);
       \node at (0pt+\x,0+\y)[draw, circle,fill=black,inner sep=1.5pt] {};
    \end{tikzpicture} \right) =0 \>,
\end{equation}
and the four-graviton amplitude at order $\tb^2$
\begin{equation}\label{eq:4pointB2}
    \begin{tikzpicture}[scale=12,baseline={([yshift=-1mm]center.base)}]
        \clip (-3pt,-3pt) rectangle (3.3pt,3pt);
        \node at (0,0)(center){};
        \node at (-2pt,-2pt) (s1){};
        \node at (-2pt,2pt)(s2){};
        \node at (2pt,2pt) (g1){};
        \node at (2pt,-2pt) (g2){};
        \node at (-2.4pt,-1.8pt) {$1^{h_1}$};
        \node at (-2.4pt,1.8pt){$2^{h_2}$};
        \node at (2.7pt,1.8pt) {$3^{h_3}$};
        \node at (2.7pt,-1.8pt) {$4^{h_4}$};

        \draw [decorate, decoration={snake, amplitude=1pt, segment length=6pt},line width=.75] (g1) -- (center);
        \draw [decorate, decoration={snake, amplitude=1pt, segment length=6pt},line width=.75]  (g2) -- (center);
        \draw [decorate, decoration={snake, amplitude=1pt, segment length=6pt},line width=.75]  (s1) -- (center);
        \draw [decorate, decoration={snake, amplitude=1pt, segment length=6pt},line width=.75]  (s2) -- (center);

        \node at (0,0) [circle, draw, fill=gray!45!white, inner sep=2pt]{\small $\partial^4$};
    \end{tikzpicture}
    \hspace{0.3cm}\overset{(\tb^2)}{=}\> \sum_{\rm perm} \> \left(
    \frac{1}{8}
    \begin{tikzpicture}[scale=12,baseline={([yshift=-1mm]center.base)}]
    \clip (-3pt,-2pt) rectangle (3pt,2pt);
    \def\x{0};
        \def\y{0};
        \node at (-1pt+\x,0+\y) (one) {};
        \node at (1pt+\x,0+\y) (two) {};
        \node at (-2.5pt+\x,2pt+\y) (oneA) {};
        \node at (-2.5pt+\x,-2pt+\y) (oneB) {};
        \node at (2.5pt+\x,2pt+\y) (twoA) {};
        \node at (2.5pt+\x,-2pt+\y) (twoB) {};

        \draw [decorate, decoration={snake, amplitude=1pt, segment length=6pt},line width=.75]  (one.center) -- (two.center);
        \draw [decorate, decoration={snake, amplitude=1pt, segment length=6pt},line width=.75]  (oneA) -- (one.center);
        \draw [decorate, decoration={snake, amplitude=1pt, segment length=6pt},line width=.75]  (oneB) -- (one.center);
        \draw [decorate, decoration={snake, amplitude=1pt, segment length=6pt},line width=.75]  (twoA) -- (two.center);
        \draw [decorate, decoration={snake, amplitude=1pt, segment length=6pt},line width=.75]  (twoB) -- (two.center);
        \node at (-1pt+\x,0+\y)[draw, circle,fill=black,inner sep=1.5pt] {};
        \node at (1pt+\x,0+\y)[draw, circle,fill=black,inner sep=1.5pt] {};
    \end{tikzpicture}
    +\frac{1}{4}
     \begin{tikzpicture}[scale=12,baseline={([yshift=-1mm]center.base)}]
    \clip (-3pt,-2pt) rectangle (3pt,2pt);
    \def\x{0};
        \def\y{0};
        \node at (-1pt+\x,0+\y) (one) {};
        \node at (1pt+\x,0+\y) (two) {};
        \node at (-2.5pt+\x,2pt+\y) (oneA) {};
        \node at (-2.5pt+\x,-2pt+\y) (oneB) {};
        \node at (2.5pt+\x,2pt+\y) (twoA) {};
        \node at (2.5pt+\x,-2pt+\y) (twoB) {};

        \draw [decorate, decoration={snake, amplitude=1pt, segment length=6pt},line width=.75]  (one.center) -- (two.center);
        \draw [decorate, decoration={snake, amplitude=1pt, segment length=6pt},line width=.75]  (oneA) -- (one.center);
        \draw [decorate, decoration={snake, amplitude=1pt, segment length=6pt},line width=.75]  (oneB) -- (one.center);
        \draw [decorate, decoration={snake, amplitude=1pt, segment length=6pt},line width=.75]  (twoA) -- (two.center);
        \draw [decorate, decoration={snake, amplitude=1pt, segment length=6pt},line width=.75]  (twoB) -- (two.center);
       \node at (-1pt+\x,0+\y)[draw, circle,fill=black,inner sep=1.5pt] {};
       \node at (0pt+\x,0+\y)[draw, circle,fill=black,inner sep=1.5pt] {};
    \end{tikzpicture}
    +\frac{1}{8}
   \begin{tikzpicture}[scale=12,baseline={([yshift=-1mm]center.base)}]
    \clip (-3pt,-2pt) rectangle (3pt,2pt);
    \def\x{0};
        \def\y{0};
        \node at (-1pt+\x,0+\y) (one) {};
        \node at (1pt+\x,0+\y) (two) {};
        \node at (-2.5pt+\x,2pt+\y) (oneA) {};
        \node at (-2.5pt+\x,-2pt+\y) (oneB) {};
        \node at (2.5pt+\x,2pt+\y) (twoA) {};
        \node at (2.5pt+\x,-2pt+\y) (twoB) {};

        \draw [decorate, decoration={snake, amplitude=1pt, segment length=6pt},line width=.75] (one.center) -- (two.center);
        \draw [decorate, decoration={snake, amplitude=1pt, segment length=6pt},line width=.75]  (oneA) -- (one.center);
        \draw [decorate, decoration={snake, amplitude=1pt, segment length=6pt},line width=.75]  (oneB) -- (one.center);
        \draw [decorate, decoration={snake, amplitude=1pt, segment length=6pt},line width=.75]  (twoA) -- (two.center);
        \draw [decorate, decoration={snake, amplitude=1pt, segment length=6pt},line width=.75]  (twoB) -- (two.center);
       \node at (-0.5pt+\x,0+\y)[draw, circle,fill=black,inner sep=1.5pt] {};
       \node at (0.5pt+\x,0+\y)[draw, circle,fill=black,inner sep=1.5pt] {};
    \end{tikzpicture} \right) =0 \>,
\end{equation}
where the sum runs over all possible permutations of the external legs and appropriate symmetry factors have been associated to each diagram. The vanishing of these amplitudes is ensured by the on-shell argument given in  Section~\ref{section:3}. To simultaneously satisfy \eqref{eq:22pointB} and~\eqref{eq:4pointB2}, one needs   
\begin{equation}\label{eq:rule1}
    \begin{tikzpicture}[scale=12,baseline={([yshift=-1mm]centro.base)}]
    \def\x{0};
    \def\y{0};
        \node at (0+\x,0+\y) (centro) {};
        \node at (2pt+\x,0+\y) (tre) {};
        \node at (2pt+\x,-0.6pt+\y) {$\mu \nu$};
        \node at (-1pt+\x,2pt+\y) (due) {};
        \node at (-1pt+\x,-2pt+\y) (uno) {};

        \draw [decorate, decoration={snake, amplitude=1pt, segment length=6pt},line width=.75] (uno) -- (centro.center);
        \draw [decorate, decoration={snake, amplitude=1pt, segment length=6pt},line width=.75] (due) -- (centro.center);
        \draw [decorate, decoration={snake, amplitude=1pt, segment length=6pt},line width=.75] (centro.center) -- (tre);
        \node at (centro) [draw,circle, fill=black, inner sep=2pt] {};

        \node at (3.5pt,0) {+};

        \def\x{6pt};
        \def\y{0};
        \node at (0+\x,0+\y) (centro) {};
        \node at (3pt+\x,0+\y) (tre) {};
        \node at (3pt+\x,-0.6pt+\y) {$\mu \nu$};
        \node at (-1pt+\x,2pt+\y) (due) {};
        \node at (-1pt+\x,-2pt+\y) (uno) {};

        \draw [decorate, decoration={snake, amplitude=1pt, segment length=6pt},line width=.75] (uno) -- (centro.center);
        \draw [decorate, decoration={snake, amplitude=1pt, segment length=6pt},line width=.75] (due) -- (centro.center);
        \draw [decorate, decoration={snake, amplitude=1pt, segment length=6pt},line width=.75] (centro.center) -- (tre);
        \node at (1.5pt+\x,0+\y) [draw,circle, fill=black, inner sep=2pt] {};
    \end{tikzpicture} \hspace{0.4cm}=0 \>,
\end{equation}
where the line carrying the Lorentz indices is off shell, whereas the other two lines are on shell. Here we recovered the diagrammatic identity  originally found by direct computation which is displayed in Figure~\ref{pallo},  from a purely on-shell argument. 

We can use this identity as a replacement rule to turn insertions of  $R^2$-type vertices into propagator insertions,  
or the other way around. Such replacements lead to great simplifications and are in some cases sufficient to prove the vanishing of entire classes of amplitudes. Examples of this situation are the four-graviton amplitude for $B>2$ and the five-graviton amplitude for $B>3$, whose vanishing is not guaranteed by the on-shell argument presented earlier; by drawing all  possible diagrams one can immediately see that the identity~\eqref{eq:rule1} implies the vanishing of these amplitudes to all orders in $\tilde{b}$.

As an example of how to obtain further diagrammatic relations from \eqref{eq:rule1} and the known vanishing amplitudes,  
consider the four-graviton amplitude at $\cO(\tb)$. 
We know that 
\begin{equation}\label{eq:4pointB1}
    \begin{tikzpicture}[scale=12,baseline={([yshift=-1mm]center.base)}]
        \clip (-3pt,-3pt) rectangle (3.3pt,3pt);
        \node at (0,0)(center){};
        \node at (-2pt,-2pt) (s1){};
        \node at (-2pt,2pt)(s2){};
        \node at (2pt,2pt) (g1){};
        \node at (2pt,-2pt) (g2){};
        \node at (-2.4pt,-1.8pt) {$1^{h_1}$};
        \node at (-2.4pt,1.8pt){$2^{h_2}$};
        \node at (2.7pt,1.8pt) {$3^{h_3}$};
        \node at (2.7pt,-1.8pt) {$4^{h_4}$};

        \draw [decorate, decoration={snake, amplitude=1pt, segment length=6pt},line width=.75] (g1) -- (center);
        \draw [decorate, decoration={snake, amplitude=1pt, segment length=6pt},line width=.75]  (g2) -- (center);
        \draw [decorate, decoration={snake, amplitude=1pt, segment length=6pt},line width=.75]  (s1) -- (center);
        \draw [decorate, decoration={snake, amplitude=1pt, segment length=6pt},line width=.75]  (s2) -- (center);

        \node at (0,0) [circle, draw, fill=gray!45!white, inner sep=2pt]{\small $\partial^4$};
    \end{tikzpicture}
    \hspace{0.3cm}\overset{(\tb)}{=}\> \sum_{\rm perm} \> \left(
    \frac{1}{8}
    \begin{tikzpicture}[scale=12,baseline={([yshift=-1mm]center.base)}]
    \clip (-3pt,-2pt) rectangle (3pt,2pt);
    \def\x{0};
        \def\y{0};
        \node at (-1pt+\x,0+\y) (one) {};
        \node at (1pt+\x,0+\y) (two) {};
        \node at (-2.5pt+\x,2pt+\y) (oneA) {};
        \node at (-2.5pt+\x,-2pt+\y) (oneB) {};
        \node at (2.5pt+\x,2pt+\y) (twoA) {};
        \node at (2.5pt+\x,-2pt+\y) (twoB) {};

        \draw [decorate, decoration={snake, amplitude=1pt, segment length=6pt},line width=.75]  (one.center) -- (two.center);
        \draw [decorate, decoration={snake, amplitude=1pt, segment length=6pt},line width=.75]  (oneA) -- (one.center);
        \draw [decorate, decoration={snake, amplitude=1pt, segment length=6pt},line width=.75]  (oneB) -- (one.center);
        \draw [decorate, decoration={snake, amplitude=1pt, segment length=6pt},line width=.75]  (twoA) -- (two.center);
        \draw [decorate, decoration={snake, amplitude=1pt, segment length=6pt},line width=.75]  (twoB) -- (two.center);
        \node at (0+\x,0+\y) [draw, circle,fill=black,inner sep=1.5pt]{};
    \end{tikzpicture}
    +\frac{1}{4}
     \begin{tikzpicture}[scale=12,baseline={([yshift=-1mm]center.base)}]
    \clip (-3pt,-2pt) rectangle (3pt,2pt);
    \def\x{0};
        \def\y{0};
        \node at (-1pt+\x,0+\y) (one) {};
        \node at (1pt+\x,0+\y) (two) {};
        \node at (-2.5pt+\x,2pt+\y) (oneA) {};
        \node at (-2.5pt+\x,-2pt+\y) (oneB) {};
        \node at (2.5pt+\x,2pt+\y) (twoA) {};
        \node at (2.5pt+\x,-2pt+\y) (twoB) {};

        \draw [decorate, decoration={snake, amplitude=1pt, segment length=6pt},line width=.75]  (one.center) -- (two.center);
        \draw [decorate, decoration={snake, amplitude=1pt, segment length=6pt},line width=.75]  (oneA) -- (one.center);
        \draw [decorate, decoration={snake, amplitude=1pt, segment length=6pt},line width=.75]  (oneB) -- (one.center);
        \draw [decorate, decoration={snake, amplitude=1pt, segment length=6pt},line width=.75]  (twoA) -- (two.center);
        \draw [decorate, decoration={snake, amplitude=1pt, segment length=6pt},line width=.75]  (twoB) -- (two.center);
       \node at (-1pt+\x,0+\y)[draw, circle,fill=black,inner sep=1.5pt] {};
    \end{tikzpicture}
    +\frac{1}{4!}
   \begin{tikzpicture}[scale=12,baseline={([yshift=-1mm]center.base)}]
    \clip (-3pt,-2pt) rectangle (2.5pt,2pt);
    \def\x{0};
        \def\y{0};
        \node at (0+\x,0+\y) (one) {};
        \node at (1pt+\x,0+\y) (two) {};
        \node at (-2pt+\x,2pt+\y) (oneA) {};
        \node at (-2pt+\x,-2pt+\y) (oneB) {};
        \node at (2pt+\x,2pt+\y) (twoA) {};
        \node at (2pt+\x,-2pt+\y) (twoB) {};

        \node at (0+\x,0+\y) [draw, circle,fill=black,inner sep=2pt]{};
        \draw [decorate, decoration={snake, amplitude=1pt, segment length=6pt},line width=.75]  (oneA) -- (one.center);
        \draw [decorate, decoration={snake, amplitude=1pt, segment length=6pt},line width=.75]  (oneB) -- (one.center);
        \draw [decorate, decoration={snake, amplitude=1pt, segment length=6pt},line width=.75]  (twoA) -- (one.center);
        \draw [decorate, decoration={snake, amplitude=1pt, segment length=6pt},line width=.75]  (twoB) -- (one.center);
    \end{tikzpicture} \right) =0 \>.
\end{equation}
Using \eqref{eq:rule1}, we obtain  a further identity: 
\begin{equation}\label{eq:rule2}
\sum_{\rm perm}\left(
    \begin{tikzpicture}[scale=12,baseline={([yshift=-1mm]centro.base)}]
    \def\x{0};
    \def\y{0};
        \node at (-1pt+\x,0+\y) (left){};
        \node at (1pt+\y,0+\y) (right){};
        \node at (-2.5pt,1.5pt+\y) (one) {};
        \node at (-2.5pt,-1.5pt+\y) (two) {};
        \node at (2.5pt+\x,1.5pt+\y) (three) {};
        \node at (2.5pt+\x,-1.5pt+\y) (four) {};

        \draw [decorate, decoration={snake, amplitude=1pt, segment length=6pt},line width=.75] (left.center) -- (right.center);
        \draw [decorate, decoration={snake, amplitude=1pt, segment length=6pt},line width=.75] (one.center) -- (left.center);
        \draw [decorate, decoration={snake, amplitude=1pt, segment length=6pt},line width=.75] (two.center) -- (left.center);
        \draw [decorate, decoration={snake, amplitude=1pt, segment length=6pt},line width=.75] (three.center) -- (right.center);
        \draw [decorate, decoration={snake, amplitude=1pt, segment length=6pt},line width=.75] (four.center) -- (right.center);
        \node at (-1pt+\x,0+\y) [draw,circle, fill=black, inner sep=2pt] {};

        \node at (4pt,0) {+};
        \node at (5pt,0) {$\dfrac{1}{3}$};

        \def\x{7.5pt};
        \def\y{0};
        \node at (0+\x,0+\y) (centro) {};
        \draw [decorate, decoration={snake, amplitude=1pt, segment length=6pt},line width=.75] (-1.5pt+\x,1.5pt+\y) -- (1.5pt+\x,-1.5pt+\y);
        \draw [decorate, decoration={snake, amplitude=1pt, segment length=6pt},line width=.75] (1.5pt+\x,1.5pt+\y) -- (-1.5pt+\x,-1.5pt+\y);
        \node at (0+\x,0+\y) [draw,circle, fill=black, inner sep=2pt] {};
    \end{tikzpicture} \hspace{0.2cm}\right) \hspace{0.4cm}=0 \>,
\end{equation}
where all external legs are on-shell.

To find a  new off-shell identity  one has to  look at  the two-scalar/three-graviton amplitude at  $\cO(\tb)$ or the six-graviton amplitude at  $\cO(\tb^2)$.
After applying~\eqref{eq:rule1} to cancel as many terms as possible, we are left with a sum of four diagrams, which we know must vanish, {\it i.e.}
\begin{equation}
\label{eq:rule3}
    \sum_{\rm perm} \> \left(
    \frac{1}{3}
    \begin{tikzpicture}[scale=12,baseline={([yshift=-1mm]center.base)}]
    \clip (-3.5pt,-2pt) rectangle (2.3pt,2pt);
    \def\x{0};
        \def\y{0};
        \node at (-1pt+\x,0+\y) (one) {};
        \node at (1pt+\x,0+\y) (two) {};
        \node at (1.5pt+\x,-0.5pt+\y){$\mu \nu$};
        \node at (-3pt+\x,0+\y) (three) {};
        \node at (-2.5pt+\x,2pt+\y) (oneA) {};
        \node at (-2.5pt+\x,-2pt+\y) (oneB) {};
        \node at (2.5pt+\x,2pt+\y) (twoA) {};
        \node at (2.5pt+\x,-2pt+\y) (twoB) {};

        \draw [decorate, decoration={snake, amplitude=1pt, segment length=6pt},line width=.75]  (one.center) -- (two.center);
        \draw [decorate, decoration={snake, amplitude=1pt, segment length=6pt},line width=.75]  (oneA) -- (one.center);
        \draw [decorate, decoration={snake, amplitude=1pt, segment length=6pt},line width=.75]  (oneB) -- (one.center);
        \draw [decorate, decoration={snake, amplitude=1pt, segment length=6pt},line width=.75] (three) -- (one.center);
        \node at (0+\x,0+\y) [draw, circle,fill=black,inner sep=1.5pt]{};
    \end{tikzpicture}
    +\frac{1}{3}
     \begin{tikzpicture}[scale=12,baseline={([yshift=-1mm]center.base)}]
    \clip (-3.3pt,-2pt) rectangle (2.3pt,2pt);
    \def\x{0};
        \def\y{0};
        \node at (-1pt+\x,0+\y) (one) {};
        \node at (1pt+\x,0+\y) (two) {};
        \node at (-3pt+\x,0+\y) (three) {};
        \node at (1.5pt+\x,-0.5pt+\y){$\mu \nu$};
        \node at (-2.5pt+\x,2pt+\y) (oneA) {};
        \node at (-2.5pt+\x,-2pt+\y) (oneB) {};
        \node at (2.5pt+\x,2pt+\y) (twoA) {};
        \node at (2.5pt+\x,-2pt+\y) (twoB) {};

        \draw [decorate, decoration={snake, amplitude=1pt, segment length=6pt},line width=.75]  (one.center) -- (two.center);
        \draw [decorate, decoration={snake, amplitude=1pt, segment length=6pt},line width=.75]  (oneA) -- (one.center);
        \draw [decorate, decoration={snake, amplitude=1pt, segment length=6pt},line width=.75]  (oneB) -- (one.center);
        \draw [decorate, decoration={snake, amplitude=1pt, segment length=6pt},line width=.75] (three) -- (one.center);
        \node at (-1pt+\x,0+\y)[draw, circle,fill=black,inner sep=1.5pt] {};
    \end{tikzpicture}
    +
     \begin{tikzpicture}[scale=12,baseline={([yshift=-1mm]center.base)}]
    \clip (-2.7pt,-2pt) rectangle (3.5pt,2pt);
    \def\x{0};
        \def\y{0};
        \node at (-1pt+\x,0+\y) (one) {};
        \node at (2pt+\x,0+\y) (two) {};
        \node at (0pt+\x,2pt+\y) (three) {};
        \node at (2.5pt+\x,-0.5pt+\y){$\mu \nu$};
        \node at (-2.5pt+\x,2pt+\y) (oneA) {};
        \node at (-2.5pt+\x,-2pt+\y) (oneB) {};
        \node at (2.5pt+\x,2pt+\y) (twoA) {};
        \node at (2.5pt+\x,-2pt+\y) (twoB) {};

        \draw [decorate, decoration={snake, amplitude=1pt, segment length=6pt},line width=.75]  (one.center) -- (two.center);
        \draw [decorate, decoration={snake, amplitude=1pt, segment length=6pt},line width=.75]  (oneA) -- (one.center);
        \draw [decorate, decoration={snake, amplitude=1pt, segment length=6pt},line width=.75]  (oneB) -- (one.center);
        \draw [decorate, decoration={snake, amplitude=1pt, segment length=6pt},line width=.75] (three) -- (0+\x,0+\y);
        \node at (0+\x,0+\y) [draw, circle,fill=black,inner sep=1.5pt]{};
    \end{tikzpicture}
    +
    \begin{tikzpicture}[scale=12,baseline={([yshift=-1mm]center.base)}]
    \clip (-2.7pt,-2pt) rectangle (3.5pt,2pt);
    \def\x{0};
        \def\y{0};
        \node at (-1pt+\x,0+\y) (one) {};
        \node at (2pt+\x,0+\y) (two) {};
        \node at (0pt+\x,2pt+\y) (three) {};
        \node at (2.5pt+\x,-0.5pt+\y){$\mu \nu$};
        \node at (-2.5pt+\x,2pt+\y) (oneA) {};
        \node at (-2.5pt+\x,-2pt+\y) (oneB) {};
        \node at (2.5pt+\x,2pt+\y) (twoA) {};
        \node at (2.5pt+\x,-2pt+\y) (twoB) {};

        \draw [decorate, decoration={snake, amplitude=1pt, segment length=6pt},line width=.75]  (one.center) -- (two.center);
        \draw [decorate, decoration={snake, amplitude=1pt, segment length=6pt},line width=.75]  (oneA) -- (one.center);
        \draw [decorate, decoration={snake, amplitude=1pt, segment length=6pt},line width=.75]  (oneB) -- (one.center);
        \draw [decorate, decoration={snake, amplitude=1pt, segment length=6pt},line width=.75] (three) -- (0+\x,0+\y);
        \node at (1pt+\x,0+\y) [draw, circle,fill=black,inner sep=1.5pt]{};
    \end{tikzpicture}
    \right) =0 \>,
\end{equation}
where the permutations are  over the on-shell legs. As a consistency check, notice that this identity reduces to~\eqref{eq:rule2} once we put  the fourth leg on-shell. 

The identity~\eqref{eq:rule3} combined with~\eqref{eq:rule1} is sufficient to prove that the corrections to the six-graviton and 
two-scalar/three-graviton amplitudes induced by the four-derivative couplings  vanish to all orders in $\tb$.
Iterating this procedure at higher multiplicity and power of the coupling $\tb$, one will find additional identities involving a higher and higher number of particles. We expect the combination of all of these identities to be sufficient to guarantee the vanishing of the $R^2$ modified $n$-graviton and two-scalar/$n$-graviton amplitudes for any $n$ and  any power of~$\tb$.

\section{Conclusions}

In summary we have shown from different but complementary angles that amplitudes contributing
to the computation of the Newtonian potential receive no corrections from curvature squared terms. As we have seen in the previous  section, this statement not only applies to linear order in the couplings $a$ and $b$, as appropriate for an effective field theory treatment, but also continues to hold  to higher orders in the couplings.
It  would be very interesting to settle this question for arbitrarily high orders, and 
we expect that  amplitude techniques may provide an alternative, more efficient method than field redefinitions, which become quickly very cumbersome at high orders in the parameters.
Similarly it would be interesting to revisit the case of terms cubic (or higher) in the curvature, and test whether there exist appropriate  field redefinitions that remove terms involving the Ricci scalar and tensor while preserving the minimal coupling of the heavy scalars (up to contact terms involving four or more scalars). Also for this case the amplitudes/on-shell techniques employed in  Section~\ref{section:3} may prove useful. Given the results of \cite{Metsaev:1986yb}, we expect a positive answer to this question, which would leave the cubic corrections computed in \cite{Brandhuber:2019qpg,Emond:2019crr} as the first higher-derivative corrections to EH gravity that can modify  the Newtonian potential.

\vspace{1cm}

\section*{Acknowledgements}

We would like to thank Tim Clifton, Rodolfo Russo, Bill Spence and Congkao Wen for  useful  discussions. This work  was supported by the Science and Technology Facilities Council (STFC) Consolidated Grant ST/P000754/1 \textit{``String theory, gauge theory \& duality''}, and by the European Union's Horizon 2020 research and innovation programme under the Marie Sk\l{}odowska-Curie grant agreement No.~764850 {\it ``\href{https://sagex.org}{SAGEX}''}.

\newpage

\appendix
\section{Feynman rules}
\label{sec::Feynmanrules}
We collect in this appendix the relevant Feynman rules used in the text. The three-point graviton vertex in EH with two on-shell legs is
\begin{align}
    \begin{split}
        \!\!
        \raisebox{-.48\height}{
            \begin{tikzpicture}[line width=1.5 pt, scale=.5]
                \draw[vector, line width = .75] (0:0) -- (0:2);
                \draw[vector, line width = .75] (0:0) -- (120:2);
                \draw[vector, line width = .75] (0:0) -- (240:2);
                \node at (-120:3) {$1$};
                \node at (120:3) {$2$};
                \node at (0:2.6) {$\alpha\beta$};
            \end{tikzpicture}
        }
        \ =& \
        i \kappa\, \epsilon_{1}\cdot\epsilon_{2}\, \bigg[ \frac{3}{2} \epsilon_{1}\cdot\epsilon_{2}\, p_{1}\cdot p_{2}\, \eta^{\alpha \beta} - \epsilon_{1}\cdot\epsilon_{2}\, p_{1}^{\alpha} p_{1}^{\beta} - \epsilon_{1}\cdot\epsilon_{2}\, p_{1}^{( \alpha} p_{2}^{\beta )} - \epsilon_{1}\cdot\epsilon_{2}\, p_{2}^{ \alpha} p_{2}^{\beta } \\[-3em]
        &- \epsilon_{1}\cdot p_{2}\, \epsilon_{2}\cdot p_{1}\, \eta^{\alpha \beta}+2 \epsilon_{1}\cdot p_{2}\, \epsilon_{2}^{( \alpha} p_{2}^{\beta )}+ 2 \epsilon_{2}\cdot p_{1}\, \epsilon_{1}^{( \alpha} p_{1}^{\beta )}-2 p_{1}\cdot p_{2}\, \epsilon_{1}^{( \alpha} \epsilon_{2}^{\beta )}\bigg]\\[.2em]
        &-i\kappa \big( \epsilon_{2}\cdot p_{1}\, \epsilon_{1}^{\alpha} - \epsilon_{1}\cdot p_{2}\, \epsilon_{2}^{\alpha} \big) \big( \epsilon_{2}\cdot p_{1}\, \epsilon_{1}^{\beta} - \epsilon_{1}\cdot p_{2}\, \epsilon_{2}^{\beta} \big)\, , 
    \end{split}
\end{align}

The Feynman rules for insertions of $R^2$ (denoted by a bullet) and $R^{\mu \nu}R_{\mu \nu}$ (denoted by a cross) are: 
\begin{align}
    \begin{split}
        \!\!
        \raisebox{-.48\height}{
            \begin{tikzpicture}[line width=1.5 pt,scale=.5]
                \draw[vector, line width = .75] (0:0) -- (0:2);
                \draw[vector, line width = .75] (0:0) -- (120:2);
                \draw[vector, line width = .75] (0:0) -- (240:2);
                \node [circle,fill=black,inner sep=2pt]at (0,0){};
                \node at (-120:3) {$1$};
                \node at (120:3) {$2$};
                \node at (0:2.6) {$\alpha\beta$};
            \end{tikzpicture}
        }
         &= \  i b\, \kappa^3 \, \epsilon_{1}\cdot\epsilon_{2}\, \big( -3\epsilon_{1}\cdot\epsilon_{2} \, p_{1}\cdot p_{2}+2 \epsilon_{1}\cdot p_{2} \epsilon_{2}\cdot p_{1} \big) \big( k^2\, \eta^{\alpha \beta} - k^{\alpha} k^{\beta} \big)\, , 
    \end{split}
    \\ \cr
    \begin{tikzpicture}[scale=15]
    \draw [vector,line width=.75](-1.5pt,0)--(1.5pt,0);
    \node at (0,0) [circle,fill=black,inner sep=1.5]{};
    \end{tikzpicture}
   & = 2i b \, \kappa^2 \, \big( k^2\, \eta^{\alpha \beta} - k^{\alpha} k^{\beta} \big)\big( k^2\, \eta^{\mu \nu} - k^{\mu} k^{\nu} \big)\, , 
\end{align}

\begin{align}
    \begin{split}
    \begin{tikzpicture}[line width=1.5 pt,scale=.5,,baseline={([yshift=-1mm]cross.base)}]
                \draw[vector, line width = .75] (0:0) -- (0:2);
                \draw[vector, line width = .75] (0:0) -- (120:2);
                \draw[vector, line width = .75] (0:0) -- (240:2);
                \node [draw, cross,inner sep=3pt,line width=1.75] at (0,0)(cross){};
                \node at (-120:3) {$1$};
                \node at (120:3) {$2$};
                \node at (0:2.6) {$\alpha\beta$};
            \end{tikzpicture}
            &= i a\, \kappa^3\, 
            \epsilon_{1}\cdot\epsilon_{2}\,\Bigg[ \frac{k^2}{2} \left( \epsilon_{1}\cdot p_{2}\,  \epsilon_{2}\cdot p_{1} - \frac{3}{4} \epsilon_{1}\cdot\epsilon_{2}\, k^2 \right) \eta^{\alpha \beta} +\frac{1}{2} \epsilon_{1}\cdot\epsilon_{2}\, k^2\, p_{1}^{( \alpha} p_{2}^{\beta )}\\[-2.3em]
            &+ \left(\frac{1}{4} \epsilon_{1}\cdot\epsilon_{2}\, k^2 - \epsilon_{1}\cdot p_{2}\, \epsilon_{2}\cdot p_{1}\right) k^{\alpha} k^{\beta}+ k^2 \left( \epsilon_{1}\cdot p_{2}\, \epsilon_{2}^{( \alpha} p_{2}^{\beta )} + \epsilon_{2}\cdot p_{1}\, \epsilon_{1}^{( \alpha} p_{1}^{\beta )} \right)\\[.3em]
            &- \frac{k^4}{2} \epsilon_{1}^{( \alpha} \epsilon_{2}^{\beta )}\Bigg]-i a\, \frac{k^2}{2} \left( \epsilon_{2}\cdot p_{1}\, \epsilon_{1}^{\alpha} - \epsilon_{1}\cdot p_{2}\, \epsilon_{2}^{\alpha} \right) \left( \epsilon_{2}\cdot p_{1}\, \epsilon_{1}^{\beta} - \epsilon_{1}\cdot p_{2}\, \epsilon_{2}^{\beta} \right)
            \, ,
        \end{split}
\\
\cr
\cr
\begin{split}
    \begin{tikzpicture}[scale=15]
    \node at (-1.6pt,0)(one) {};
    \node at (1.6pt,0)(two) {};
    \draw [vector,line width=.75](one.center) -- (0,0);
    \draw [vector,line width=.75](two.center) -- (0,0);
    \node at (0,0)[draw, cross,inner sep=3pt,line width=1.75]{};
    \end{tikzpicture}
  &  = i a \, \kappa^2 \, {\rm Sym}\!\left[ k^{\alpha} k^{\beta} k^{\mu} k^{\nu} - \frac{1}{2}\eta^{\alpha \beta} k^{\mu} k^{\nu} k^2 -\eta^{\alpha \mu} k^{\beta} k^{\nu} k^2 +\frac{1}{2}\eta^{\mu \nu} \eta^{\alpha \beta} k^4+\frac{1}{2}\eta^{\alpha \mu} \eta^{\beta \nu} k^4 \right] ,
 \\ &
 \\ &
\end{split}
\end{align}
where the symmetrisation is both over the indices $\mu \longleftrightarrow \nu$, $\alpha \longleftrightarrow \beta$ and $(\mu ,\nu) \longleftrightarrow (\alpha ,\beta)$.

\vspace{0.7cm}

\appendix

\bibliographystyle{utphys}
\bibliography{remainder}

\providecommand{\href}[2]{#2}\begingroup\raggedright\begin{thebibliography}{10}

\bibitem{Clifton:2011jh}
T.~Clifton, P.~G. Ferreira, A.~Padilla, and C.~Skordis, ``{Modified Gravity and
  Cosmology},'' \href{http://dx.doi.org/10.1016/j.physrep.2012.01.001}{{\em
  Phys. Rept.} {\bfseries 513} (2012) 1--189},
\href{http://arxiv.org/abs/1106.2476}{{\ttfamily arXiv:1106.2476
  [astro-ph.CO]}}.

\bibitem{Berti:2015itd}
E.~Berti {\em et~al.}, ``{Testing General Relativity with Present and Future
  Astrophysical Observations},''
  \href{http://dx.doi.org/10.1088/0264-9381/32/24/243001}{{\em Class. Quant.
  Grav.} {\bfseries 32} (2015) 243001},
\href{http://arxiv.org/abs/1501.07274}{{\ttfamily arXiv:1501.07274 [gr-qc]}}.

\bibitem{Tseytlin:1986zz}
A.~A. Tseytlin, ``{Ambiguity in the Effective Action in String Theories},''
\href{http://dx.doi.org/10.1016/0370-2693(86)90930-5}{{\em Phys. Lett.}
  {\bfseries B176} (1986) 92--98}.

\bibitem{Deser:1986xr}
S.~Deser and A.~N. Redlich, ``{String Induced Gravity and Ghost Freedom},''
  \href{http://dx.doi.org/10.1016/0370-2693(86)90177-2}{{\em Phys. Lett.}
  {\bfseries B176} (1986) 350}.
[Erratum: Phys. Lett.B186,461(1987)].

\bibitem{Tseytlin:1986ti}
A.~A. Tseytlin, ``{Vector Field Effective Action in the Open Superstring
  Theory},'' \href{http://dx.doi.org/10.1016/0550-3213(86)90303-2,
  10.1016/0550-3213(87)90500-1}{{\em Nucl. Phys.} {\bfseries B276} (1986) 391}.
[Erratum: Nucl. Phys.B291,876(1987)].

\bibitem{Metsaev:1986yb}
R.~R. Metsaev and A.~A. Tseytlin, ``{Curvature Cubed Terms in String Theory
  Effective Actions},''
\href{http://dx.doi.org/10.1016/0370-2693(87)91527-9}{{\em Phys. Lett.}
  {\bfseries B185} (1987) 52--58}.

\bibitem{tHooft:1974toh}
G.~'t~Hooft and M.~J.~G. Veltman, ``{One loop divergencies in the theory of
  gravitation},''
{\em Ann. Inst. H. Poincare Phys. Theor.} {\bfseries A20} (1974) 69--94.

\bibitem{Goroff:1985sz}
M.~H. Goroff and A.~Sagnotti, ``{Quantum gravity at two loops},''
\href{http://dx.doi.org/10.1016/0370-2693(85)91470-4}{{\em Phys. Lett.}
  {\bfseries 160B} (1985) 81--86}.

\bibitem{Goroff:1985th}
M.~H. Goroff and A.~Sagnotti, ``{The Ultraviolet Behavior of Einstein
  Gravity},''
\href{http://dx.doi.org/10.1016/0550-3213(86)90193-8}{{\em Nucl. Phys.}
  {\bfseries B266} (1986) 709--736}.

\bibitem{Bern:2015xsa}
Z.~Bern, C.~Cheung, H.-H. Chi, S.~Davies, L.~Dixon, and J.~Nohle, ``{Evanescent
  Effects Can Alter Ultraviolet Divergences in Quantum Gravity without Physical
  Consequences},'' \href{http://dx.doi.org/10.1103/PhysRevLett.115.211301}{{\em
  Phys. Rev. Lett.} {\bfseries 115} no.~21, (2015) 211301},
\href{http://arxiv.org/abs/1507.06118}{{\ttfamily arXiv:1507.06118 [hep-th]}}.

\bibitem{Bern:2017tuc}
Z.~Bern, A.~Edison, D.~Kosower, and J.~Parra-Martinez, ``{Curvature-squared
  multiplets, evanescent effects, and the U(1) anomaly in $N=4$
  supergravity},'' \href{http://dx.doi.org/10.1103/PhysRevD.96.066004}{{\em
  Phys. Rev.} {\bfseries D96} no.~6, (2017) 066004},
\href{http://arxiv.org/abs/1706.01486}{{\ttfamily arXiv:1706.01486 [hep-th]}}.

\bibitem{Iwasaki:1971vb}
Y.~Iwasaki, ``{Quantum theory of gravitation vs. classical theory -
  fourth-order potential},''
\href{http://dx.doi.org/10.1143/PTP.46.1587}{{\em Prog. Theor. Phys.}
  {\bfseries 46} (1971) 1587--1609}.

\bibitem{Iwasaki:1971iy}
Y.~Iwasaki, ``{Fourth-order gravitational potential based on quantum field
  theory},'' \href{http://dx.doi.org/10.1007/BF02770190}{{\em Lett. Nuovo Cim.}
  {\bfseries 1S2} (1971) 783--786}.
[Lett. Nuovo Cim.1,783(1971)].

\bibitem{Donoghue:1994dn}
J.~F. Donoghue, ``{General relativity as an effective field theory: The leading
  quantum corrections},''
  \href{http://dx.doi.org/10.1103/PhysRevD.50.3874}{{\em Phys. Rev.} {\bfseries
  D50} (1994) 3874--3888},
\href{http://arxiv.org/abs/gr-qc/9405057}{{\ttfamily arXiv:gr-qc/9405057
  [gr-qc]}}.

\bibitem{BjerrumBohr:2002kt}
N.~E.~J. Bjerrum-Bohr, J.~F. Donoghue, and B.~R. Holstein, ``{Quantum
  gravitational corrections to the nonrelativistic scattering potential of two
  masses},'' \href{http://dx.doi.org/10.1103/PhysRevD.71.069903,
  10.1103/PhysRevD.67.084033}{{\em Phys. Rev.} {\bfseries D67} (2003) 084033},
  \href{http://arxiv.org/abs/hep-th/0211072}{{\ttfamily arXiv:hep-th/0211072
  [hep-th]}}.
[Erratum: Phys. Rev.D71,069903(2005)].

\bibitem{Neill:2013wsa}
D.~Neill and I.~Z. Rothstein, ``{Classical Space-Times from the S Matrix},''
  \href{http://dx.doi.org/10.1016/j.nuclphysb.2013.09.007}{{\em Nucl. Phys.}
  {\bfseries B877} (2013) 177--189},
\href{http://arxiv.org/abs/1304.7263}{{\ttfamily arXiv:1304.7263 [hep-th]}}.

\bibitem{Bjerrum-Bohr:2013bxa}
N.~E.~J. Bjerrum-Bohr, J.~F. Donoghue, and P.~Vanhove, ``{On-shell Techniques
  and Universal Results in Quantum Gravity},''
  \href{http://dx.doi.org/10.1007/JHEP02(2014)111}{{\em JHEP} {\bfseries 02}
  (2014) 111},
\href{http://arxiv.org/abs/1309.0804}{{\ttfamily arXiv:1309.0804 [hep-th]}}.

\bibitem{Brandhuber:2019qpg}
A.~Brandhuber and G.~Travaglini, ``{On higher-derivative effects on the
  gravitational potential and particle bending},''
\href{http://arxiv.org/abs/1905.05657}{{\ttfamily arXiv:1905.05657 [hep-th]}}.

\bibitem{Emond:2019crr}
W.~T. Emond and N.~Moynihan, ``{Scattering Amplitudes, Black Holes and Leading
  Singularities in Cubic Theories of Gravity},''
\href{http://arxiv.org/abs/1905.08213}{{\ttfamily arXiv:1905.08213 [hep-th]}}.

\bibitem{Cristofoli:2019ewu}
A.~Cristofoli, ``{Post-Minkowskian Hamiltonians in Modified Theories of
  Gravity},''
\href{http://arxiv.org/abs/1906.05209}{{\ttfamily arXiv:1906.05209 [hep-th]}}.

\bibitem{Bjerrum-Bohr:2019kec}
N.~E.~J. Bjerrum-Bohr, A.~Cristofoli, and P.~H. Damgaard, ``{Post-Minkowskian
  Scattering Angle in Einstein Gravity},''
\href{http://arxiv.org/abs/1910.09366}{{\ttfamily arXiv:1910.09366 [hep-th]}}.

\bibitem{Stelle:1976gc}
K.~S. Stelle, ``{Renormalization of Higher Derivative Quantum Gravity},''
\href{http://dx.doi.org/10.1103/PhysRevD.16.953}{{\em Phys. Rev.} {\bfseries
  D16} (1977) 953--969}.

\bibitem{Stelle:1977ry}
K.~S. Stelle, ``{Classical Gravity with Higher Derivatives},''
\href{http://dx.doi.org/10.1007/BF00760427}{{\em Gen. Rel. Grav.} {\bfseries 9}
  (1978) 353--371}.

\bibitem{Julve:1978xn}
J.~Julve and M.~Tonin, ``{Quantum Gravity with Higher Derivative Terms},''
\href{http://dx.doi.org/10.1007/BF02748637}{{\em Nuovo Cim.} {\bfseries B46}
  (1978) 137--152}.

\bibitem{Alvarez-Gaume:2015rwa}
L.~Alvarez-Gaume, A.~Kehagias, C.~Kounnas, D.~Lüst, and A.~Riotto, ``{Aspects
  of Quadratic Gravity},'' \href{http://dx.doi.org/10.1002/prop.201500100}{{\em
  Fortsch. Phys.} {\bfseries 64} no.~2-3, (2016) 176--189},
\href{http://arxiv.org/abs/1505.07657}{{\ttfamily arXiv:1505.07657 [hep-th]}}.

\bibitem{Simon:1990ic}
J.~Z. Simon, ``{Higher Derivative Lagrangians, Nonlocality, Problems and
  Solutions},''
\href{http://dx.doi.org/10.1103/PhysRevD.41.3720}{{\em Phys. Rev.} {\bfseries
  D41} (1990) 3720}.

\bibitem{Simon:1990jn}
J.~Z. Simon, ``{The Stability of flat space, semiclassical gravity, and higher
  derivatives},''
\href{http://dx.doi.org/10.1103/PhysRevD.43.3308}{{\em Phys. Rev.} {\bfseries
  D43} (1991) 3308--3316}.

\bibitem{Donoghue:1995cz}
J.~F. Donoghue, ``{Introduction to the effective field theory description of
  gravity},'' in {\em {Advanced School on Effective Theories Almunecar, Spain,
  June 25-July 1, 1995}}.
\newblock 1995.
\newblock
\href{http://arxiv.org/abs/gr-qc/9512024}{{\ttfamily arXiv:gr-qc/9512024
  [gr-qc]}}.
\newblock

\bibitem{Holstein:2004dn}
B.~R. Holstein and J.~F. Donoghue, ``{Classical physics and quantum loops},''
  \href{http://dx.doi.org/10.1103/PhysRevLett.93.201602}{{\em Phys. Rev. Lett.}
  {\bfseries 93} (2004) 201602},
\href{http://arxiv.org/abs/hep-th/0405239}{{\ttfamily arXiv:hep-th/0405239
  [hep-th]}}.

\bibitem{Bern:1994cg}
Z.~Bern, L.~J. Dixon, D.~C. Dunbar, and D.~A. Kosower, ``{Fusing gauge theory
  tree amplitudes into loop amplitudes},''
  \href{http://dx.doi.org/10.1016/0550-3213(94)00488-Z}{{\em Nucl. Phys.}
  {\bfseries B435} (1995) 59--101},
\href{http://arxiv.org/abs/hep-ph/9409265}{{\ttfamily arXiv:hep-ph/9409265
  [hep-ph]}}.

\bibitem{Bern:1994zx}
Z.~Bern, L.~J. Dixon, D.~C. Dunbar, and D.~A. Kosower, ``{One loop $n$-point
  gauge theory amplitudes, unitarity and collinear limits},''
  \href{http://dx.doi.org/10.1016/0550-3213(94)90179-1}{{\em Nucl. Phys.}
  {\bfseries B425} (1994) 217--260},
\href{http://arxiv.org/abs/hep-ph/9403226}{{\ttfamily arXiv:hep-ph/9403226
  [hep-ph]}}.

\bibitem{Bjerrum-Bohr:2014zsa}
N.~E.~J. Bjerrum-Bohr, J.~F. Donoghue, B.~R. Holstein, L.~Planté, and
  P.~Vanhove, ``{Bending of Light in Quantum Gravity},''
  \href{http://dx.doi.org/10.1103/PhysRevLett.114.061301}{{\em Phys. Rev.
  Lett.} {\bfseries 114} no.~6, (2015) 061301},
\href{http://arxiv.org/abs/1410.7590}{{\ttfamily arXiv:1410.7590 [hep-th]}}.

\bibitem{Bjerrum-Bohr:2016hpa}
N.~E.~J. Bjerrum-Bohr, J.~F. Donoghue, B.~R. Holstein, L.~Plante, and
  P.~Vanhove, ``{Light-like Scattering in Quantum Gravity},''
  \href{http://dx.doi.org/10.1007/JHEP11(2016)117}{{\em JHEP} {\bfseries 11}
  (2016) 117},
\href{http://arxiv.org/abs/1609.07477}{{\ttfamily arXiv:1609.07477 [hep-th]}}.

\bibitem{Bai:2016ivl}
D.~Bai and Y.~Huang, ``{More on the Bending of Light in Quantum Gravity},''
  \href{http://dx.doi.org/10.1103/PhysRevD.95.064045}{{\em Phys. Rev.}
  {\bfseries D95} no.~6, (2017) 064045},
\href{http://arxiv.org/abs/1612.07629}{{\ttfamily arXiv:1612.07629 [hep-th]}}.

\bibitem{Chi:2019owc}
H.-H. Chi, ``{Graviton Bending in Quantum Gravity from One-Loop Amplitudes},''
\href{http://arxiv.org/abs/1903.07944}{{\ttfamily arXiv:1903.07944 [hep-th]}}.

\bibitem{Bern:2019nnu}
Z.~Bern, C.~Cheung, R.~Roiban, C.-H. Shen, M.~P. Solon, and M.~Zeng,
  ``{Scattering Amplitudes and the Conservative Hamiltonian for Binary Systems
  at Third Post-Minkowskian Order},''
\href{http://arxiv.org/abs/1901.04424}{{\ttfamily arXiv:1901.04424 [hep-th]}}.

\bibitem{Bern:2019crd}
Z.~Bern, C.~Cheung, R.~Roiban, C.-H. Shen, M.~P. Solon, and M.~Zeng, ``{Black
  Hole Binary Dynamics from the Double Copy and Effective Theory},''
\href{http://arxiv.org/abs/1908.01493}{{\ttfamily arXiv:1908.01493 [hep-th]}}.

\bibitem{Borchers:1962cbs}
H.~J. Borchers, ``{On structure of the algebra of field operators},''
\href{http://dx.doi.org/10.1007/BF02745645}{{\em Nuovo Cim.} {\bfseries 24}
  no.~2, (1962) 214--236}.

\bibitem{Coleman:1967vs}
S.~Coleman, ``{Soft pions},'' in {\em {Hadrons and their Interactions: Current
  and Field Algebra, Soft Pions, Supermultiplets, and Related Topics}},
  pp.~9--50.
\newblock
1967.
\newblock

\bibitem{Coleman:1969sm}
S.~R. Coleman, J.~Wess, and B.~Zumino, ``{Structure of phenomenological
  Lagrangians. 1.},''
\href{http://dx.doi.org/10.1103/PhysRev.177.2239}{{\em Phys. Rev.} {\bfseries
  177} (1969) 2239--2247}.

\bibitem{Bergere:1975tr}
M.~C. Bergere and Y.-M.~P. Lam, ``{Equivalence Theorem and Faddeev-Popov
  Ghosts},''
\href{http://dx.doi.org/10.1103/PhysRevD.13.3247}{{\em Phys. Rev.} {\bfseries
  D13} (1976) 3247--3255}.

\bibitem{Haag:1992hx}
R.~Haag, {\em {Local quantum physics: Fields, particles, algebras}}.
\newblock
1992.
\newblock

\bibitem{Ettle:2006bw}
J.~H. Ettle and T.~R. Morris, ``{Structure of the MHV-rules Lagrangian},''
  \href{http://dx.doi.org/10.1088/1126-6708/2006/08/003}{{\em JHEP} {\bfseries
  08} (2006) 003},
\href{http://arxiv.org/abs/hep-th/0605121}{{\ttfamily arXiv:hep-th/0605121
  [hep-th]}}.

\bibitem{Brandhuber:2006bf}
A.~Brandhuber, B.~Spence, and G.~Travaglini, ``{Amplitudes in Pure Yang-Mills
  and MHV Diagrams},''
  \href{http://dx.doi.org/10.1088/1126-6708/2007/02/088}{{\em JHEP} {\bfseries
  02} (2007) 088},
\href{http://arxiv.org/abs/hep-th/0612007}{{\ttfamily arXiv:hep-th/0612007
  [hep-th]}}.

\end{thebibliography}\endgroup

\end{document}